\documentclass[structabstract]{aa}
\def\apjl{ApJ}%
\def\apjs{ApJS}%
\def\ao{Appl.~Opt.}%
\def\aap{A\&A}%
\def\s4g{S$^4$G}

\usepackage{graphicx}
\usepackage{txfonts}
\begin{document} 
\title{Substellar and low-mass dwarf identification with near-infrared imaging space observatories}

   \author{B.W. Holwerda\inst{1}
          \and
          J. S. Bridge\inst{1}, 
		  R. Ryan\inst{2}, 
		  M. A. Kenworthy\inst{3},
		  N. Pirzkal\inst{2},  
		  M. Andersen\inst{4}, %
		  S. Wilkins\inst{5},
		  R. Smit\inst{6},
		  S. R. Bernard\inst{7,8},
		  T. Meshkat\inst{9}, 
          R. Steele\inst{1} \and
		  R. C. Bouwens\inst{3}
          }
          
   \institute{Department of Physics and Astronomy, 102 Natural Science Building, University of Louisville, Louisville KY 40292, USA, \email{benne.holwerda@louisville.edu}
           	\and
			Space Telescope Science Institute, 3700 San Martin Drive, Baltimore, MD 21218, USA
            \and 
            Leiden Observatory, University of Leiden, Niels Bohrweg 2, NL-2333 CA Leiden, The Netherlands
            \and
            Gemini Observatory, Southern Operations Center, c/o AURA, Casilla 603, La Serena, Chile
            \and 
            Department of Physics and Astronomy, University of Sussex, Brighton, BN1 9QH, Sussex, United Kingdom
            \and 
            Kavli Institute of Cosmology, c/o Institute of Astronomy, Madingley Road, Cambridge CB3 0HA, United Kingdom
            \and 
            School of Physics, The University of Melbourne, VIC 3010 Australia
            \and 
            ARC Centre of Excellence for All-sky Astrophysics (CAASTRO), Australia
            \and 
            IPAC, California Institute of Technology, Pasadena, CA 91125, USA}

   \date{Received September 15, 1996; accepted March 16, 1997}

 
  \abstract
   {}
   {We aim to evaluate the near-infrared colors of brown dwarfs as observed with four major infrared imaging space observatories: the Hubble Space Telescope (\emph{HST}), the James Webb Space Telescope (\emph{JWST}), the \emph{Euclid} mission, and the \emph{WFIRST} telescope. }
   {We used the {\sc SPLAT} SPEX/ISPEX spectroscopic library to map out the colors of the M-, L-, and T-type dwarfs. We have identified which color-color combination is optimal for identifying broad type and which single color is optimal to then identify the subtype (e.g., T0-9). We evaluated each observatory separately as well as the narrow-field (\emph{HST} and \emph{JWST}) and wide-field (\emph{Euclid} and \emph{WFIRST}) combinations.}
   {The \emph{Euclid} filters perform poorly typing brown dwarfs and \emph{WFIRST} performs only marginally better, despite a wider selection of filters. \emph{WFIRST}'s W146 and F062 combined with \emph{Euclid}'s Y-band discriminates somewhat better between broad brown dwarf categories. However, subtyping with any combination of \emph{Euclid} and \emph{WFIRST} observations remains uncertain due to the lack of medium or narrow-band filters. We argue that a medium band added to the \emph{WFIRST} filter selection would greatly improve its ability to preselect brown dwarfs its imaging surveys.}
   {   The \emph{HST} filters used in high-redshift searches are close to optimal to identify broad stellar type. However, the addition of F127M to the commonly used broad filter sets would allow for unambiguous subtyping. An improvement over HST is one of two broad and medium filter combinations on \emph{JWST}: pairing F140M with either F150W or F162M discriminates very well between subtypes.
}


   \maketitle
%

\section{Introduction}

Several near-infrared space telescope missions are planned in the coming decade; the \emph{James Webb Space Telescope} (\emph{JWST}), the \emph{Euclid} mission, and the \emph{Wide-field Infrared Survey Telescope} (\emph{WFIRST}). Together with the Wide Field Camera 3 (WFC3) on board the \emph{Hubble Space Telescope} (\emph{HST}), 
they represent an opportunity to study the structure of the Milky Way through the distribution of red and brown dwarf stars \citep[M-, L-, T-, and Y-type;][]{Kirkpatrick05, Burgasser06b, Cruz07, Kirkpatrick11, Dieterich14, Tinney14}. All of these observatories will survey extragalactic fields that will overlap with observations of the disk and halo components of the Milky Way that include these stellar and substellar objects.

In this respect, the \emph{HST}/WFC3 observations of extragalactic fields \citep[e.g., CANDELS and BoRG;][]{Grogin11,Koekemoer11,Trenti10} are a guide into how one could use the stars found in extragalactic fields to map our own Milky Way. This by-product of the extragalactic surveys has been explored already, often combining near-infrared images with grism spectra or proper motion data to positively identify brown dwarfs. 

For example, \cite{Pirzkal05} found M-type brown dwarfs in the Hubble Ultra Deep Field \citep[HUDF;][]{Beckwith06}, and \cite{Ryan05} identified L- and T-type dwarfs in a small set of Advanced Camera for Surveys (ACS) parallel observations. These studies primarily measure the scale height of the Milky Way disk. 

\cite{Pirzkal09} mapped M-type dwarfs in the Great Observatories Origins 
Deep Survey fields \citep[GOODS;][]{goods} using the grism observations from PEARS \cite[Probing Evolution And Reionization Spectroscopically;][]{Straughn09} for positive identification. Similarly, \cite{Ryan11} and \cite{Holwerda14} find T-type and M-/L-type dwarfs, respectively, in pure-parallel\footnote{A special mode of imaging observations offered on \emph{HST}: while the main observation is performed with the COS spectrograph, the WFC3 or ACS camera takes undithered imaging. Image quality and exposure time are consequently more limited than targeted observations.} WFC3 observations. With the increasing number of sight-lines, primarily a result of the pure-parallel campaigns of the Brightest of Reionizing Galaxies \citep[BoRG;][]{Trenti11, Calvi16, Bernard16}, statistics have improved to a point where one can model more than just the scale height of the Galactic thin disk of the Galaxy. One can model simultaneously the thickness of the disk and the  the stellar halo \citep[][]{van-Vledder16}. A second goal of theses studies is to accurately map the scale height as a function of substellar dwarf subtype. scale height as a function of subtype links the relation between Galactic wide star-formation history, the dynamical heating of the substellar population, and the cooling of the substellar dwarfs over time \citep{Ryan17}. They predict a thicker Galactic disk for later subtypes with the slope of the increase primarily depending on the Galactic star-formation history. 

A substantial motivation for these studies is to exclude Galactic brown dwarfs from the selections of high-redshift objects, which they resemble in near-infrared color space \citep[][Holwerda et al. {\em in prep}]{Caballero08, Wilkins14} and in direct imaging surveys in search of extrasolar planets. Because brown dwarfs have been found in observations that were not specifically designed for their discovery, we explore the filter sets on current and future near-infrared observatories to identify which combination of filters is optimal to identify brown dwarfs and their (sub)types.

It is reasonable to expect that of the  multitude of red or brown dwarfs still to be discovered, a great many will be found in high Galactic latitude surveys designed to observe the high redshift Universe. Previous work on these objects relied on all-sky surveys (e.g., ALLWISE and 2MASS) with (for extragalactic searches) shallow limits. A much larger volume will be probed by the deep extra-planar extragalactic surveys planned for \emph{Euclid} and \emph{WFIRST}.

Our aim is to predict which survey will produce the best brown dwarf samples and to make filter choice recommendations for surveys to improve brown dwarf identification. Our approach is largely led by our experiences with the BoRG survey \citep{Ryan11,Holwerda14} where morphologically identified stars (fully unresolved), a near-infrared color-color selection identified broad brown dwarf types, and a different single color was used to subtype a selection of these brown dwarfs. Our overall assumption in the following is that grism or spectroscopic observations of these brown dwarfs is considered an undesirable or impractical  outcome by the high redshift survey teams and that this information is therefore not available, or alternatively, the imaging mode is secondary (parallel observations), making grism observations impractical (e.g., \emph{WFIRST} guiding for coronagraphic or integral field unit observations). 

An astrophysical caveat for this paper is that we assume that all the red and brown dwarfs observed are single stars, even though it is well established that anywhere from 0-50\% of all these stars are actually in binaries, depending on type and environment \citep[e.g.,][and references therein]{Joergens03a, Burgasser06c, Ahmic07a,Dupuy12, Ward-Duong15, Opitz16, Shan17}. 
Most of the later brown dwarfs (L and T) have unique spectral signatures \citep{Burgasser16,Burgasser17a,Theissen18,Bardalez-Gagliuffi18}, causing their colors (especially medium or narrow filters) to be unique. How exactly these change as such stars are in close binaries depends on the mix of stellar types but the range of all {\sc splat} sources in comparison to just the standard star relations should give some indication. 
This assumption of no binaries has been made in all the photometric searches for red and brown dwarf stars thus far because simple photometric typing cannot distinguish well enough between close binaries of different types and single red or brown dwarfs. One can overcome this problem by using methods such as Markov Chain Monte Carlo techniques \citep[MCMC; see][]{van-Vledder16}, which allow for a fraction of the data to be erroneous. 

We divide the current and future surveys into narrow-field (\emph{HST} and \emph{JWST}) and wide-field (\emph{WFIRST} and \emph{Euclid}), and combine missions and instruments to search for observations that would be ideal to identify brown dwarfs throughout the disk and halo of our Milky Way.
Our goal for this paper is to evaluate filter combinations to separate out broad brown dwarf types (M, L, and T) and colors to subtype each brown dwarf type (e.g., M0-M9 or T0-T5). Our concern here is to explore what will constitute the minimum filter combination to type brown dwarfs reliably. 
This paper is organized as follows: 
section \ref{s:splat} summarizes the {\sc SPLAT} library used to evaluate the filters,
section \ref{s:obs} summarizes the observatories we consider here,
section \ref{s:2col} presents the results of broadly categorizing brown dwarfs in color-color space, and 
section \ref{s:1col} discusses how well any single color could be used to subtype those objects identified as bona-fide brown dwarfs.
We discuss the results in section \ref{s:discussion} with our concluding remarks in section \ref{s:conclusions}.

\begin{figure}
\includegraphics[width=0.5\textwidth]{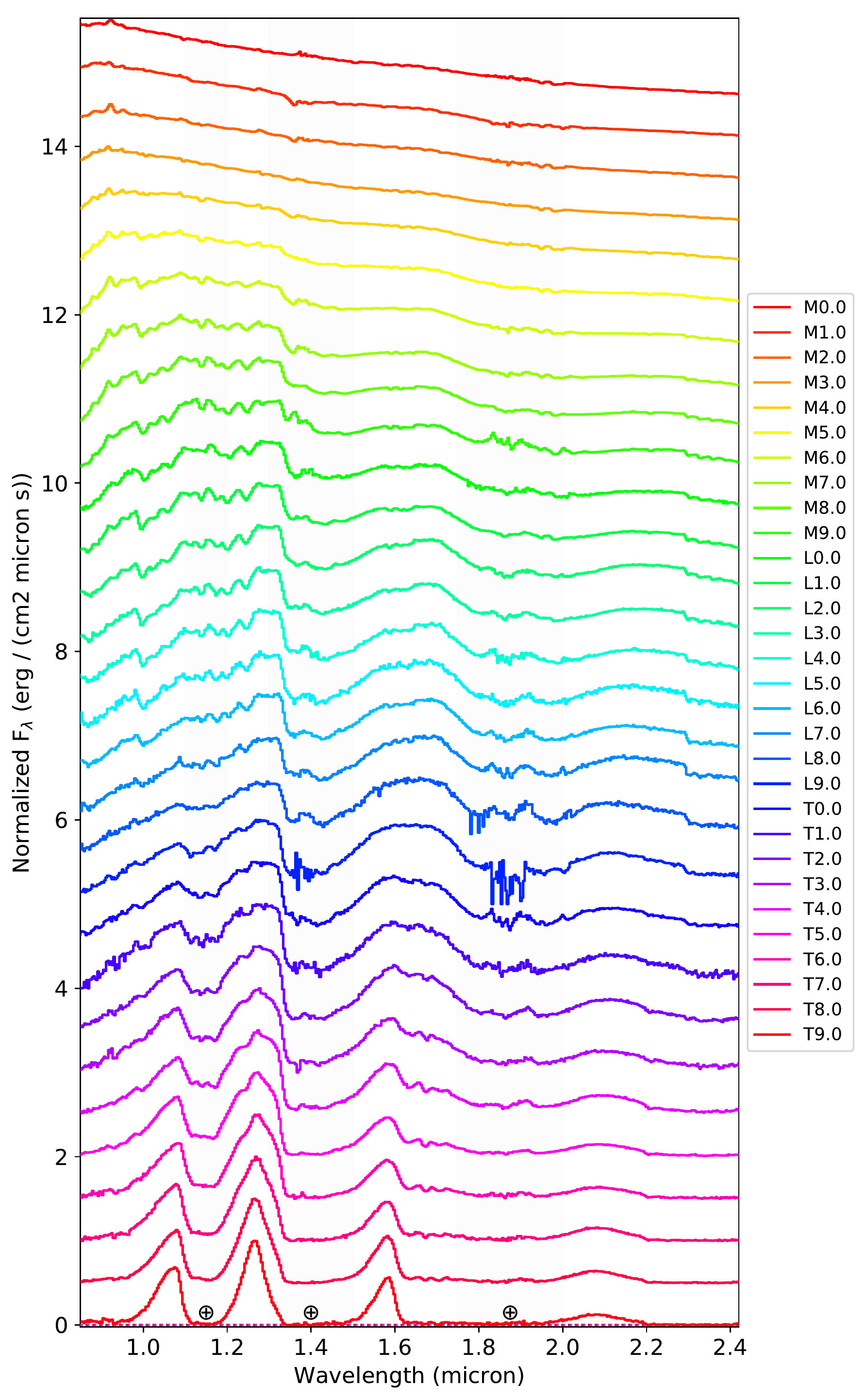}
\caption{\label{f:standards} Spectra of the standard stars in \protect{\sc SPLAT}, defining each subtype of brown dwarf from M0 to T9 for the 0.9--2.4 $\mu$m wavelength range. \protect{\sc SPLAT} contains spectral standards for dwarf classes M0 through T9, drawn from \protect\cite{Burgasser06,Kirkpatrick10}, and \protect\cite{Cushing11}.
Near-infrared colors are strongly influenced by the strength of the absorption features at 1.2 $\mu$m (methane) and 1.4 $\mu$m (water), especially in the later brown dwarf types (T and later). }
\end{figure}

\section{Red and brown dwarf spectra}
\label{s:splat}

To map the near-infrared colors of brown dwarfs, we use the {\sc Python} module {\sc SPLAT} \footnote{\protect\url{http://pono.ucsd.edu/~adam/browndwarfs/splat/}} \citep{Burgasser17}. This module contains an online repository of 1701 low-resolution, near-infrared spectra of low-temperature stars and brown dwarfs. It is built on common python packages such as {\sc Astropy, Matplotlib, NumPy, pandas} and {\sc SciPy}. We introduced the \emph{HST}, \emph{JWST}, \emph{WFIRST}, and \emph{Euclid} filters into this package using the built-in ``custom" filter option for spectrophotometry.  All colors reported are derived from Vega magnitudes. Two example tables are shown in Tables \ref{t:hst:all:mag} and \ref{t:hst:std:mag} for the all and the standard stars (defining the type) in the various HST filters. Full tables for all four observatories are available online with this publication.

We use the ensemble of spectra to map out the spectrophotometry using the built-in modules to compute the colors of near-infrared filter combinations. The built-in standard star library (Figure \ref{f:standards}) as well as the full spectral library are both used in the following work. The spectral library and classifications come from \cite{Reid01}, \cite{Testi01}, \cite{Allers07}, and \cite{Burgasser07}. 

These spectra may not be representative of the more distant red or brown dwarf stars in the halo or thick disk ; higher surface gravity, lower metallicity, and changes in the NIR absorption features as the C, N, O abundances change.

We used the standard stars from \cite{Burgasser06}, \cite{Kirkpatrick10}, and \protect\cite{Cushing11} to explore the ``ideal" color or color-color track. Using the full {\sc SPLAT} database, we illustrate the noise and variance that can be expected from a population of brown dwarfs.
Red and brown dwarf type are assigned a numerical type with the convention 0-1 (M-types), 1-2 (L-types), 2-3 (T-types) and 3 and above (Y-types), with decimal values denoting the subtype. 

\begin{figure*}
\includegraphics[width=\textwidth]{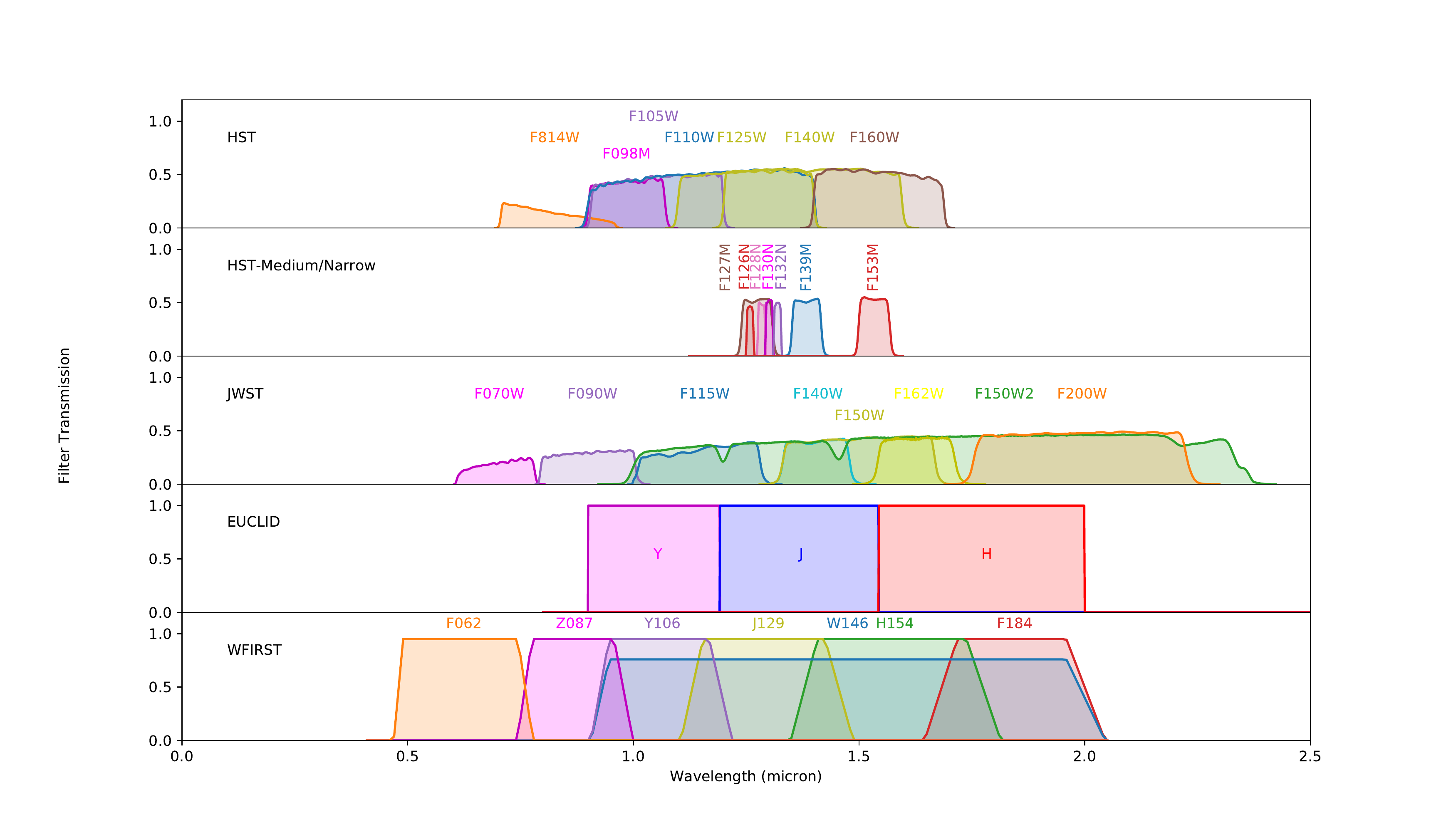}
\caption{\label{f:filters}Overview of all the near-infrared filters in the 0.5--2 $\mu$m range for the four space observatories. The top panel shows the typical filters available in high-redshift surveys with \emph{Hubble}. The second panel shows the medium and narrow bands available in \emph{HST}/WFC3 but not generally used in extragalactic surveys.
The third panel shows the \emph{JWST}/NIRCam filter sets available. The fourth row are the specified NIR filter requirements for the \emph{Euclid} mission camera. The fifth row is the current \emph{WFIRST} filter set considered for the mission. }
\end{figure*}

\section{Observatories}
\label{s:obs}

We evaluate the near-infrared filter sets of four space-borne observatories: \emph{HST} and \emph{JWST} are pointed, narrow-field observatories, while \emph{Euclid} and \emph{WFIRST} are wide-field observatories. Therefore we also evaluate the filter combinations between the first two and latter two observatories. The near-infrared filters of all four observatories are summarized in Figure \ref{f:filters}. The \emph{WFIRST} and \emph{Euclid} filters are based on mission specifications and measured prototype filters, while the \emph{HST} and \emph{JWST} filter transmission curves are measured from flight hardware.
We divide the \emph{HST} medium- and narrow-band filters in two categories, depending on whether or not they are commonly used in extragalactic surveys. One group contains mostly broad filters and the F105M filter that are used for extragalactic surveys, and other encompasses the remaining medium- and narrow-band filters. 
All four observatories have comparable point spread function (PSF) sizes across this wavelength regime (\textit{FWHM} $\sim 0.1$ arcseconds) so for our purposes, we can assume that to first order, all four observatories will perform equally well in distinguishing stars from galaxies in the near-infrared imaging. The problem is therefore reduced to how well stellar objects can be typed by one or more near-infrared colors. 

\subsection{\emph{Hubble Space Telescope} Wide Field Camera 3}

The Wide Field Camera 3 \citep[WFC3;][]{Kimble08} added unique ultraviolet and near-infrared capabilities to the \emph{HST} when it was installed. We consider the typical set of filters on WFC3 that are used for extragalactic observations: F098M, F105W, F125W, F140M, and F160W. 
We also include the near-infrared/optical filter F814W in this selection as it is typically available in high-redshift surveys such as CANDELS \citep{Grogin11,Koekemoer11}.

\subsection{\emph{Hubble Space Telescope} Wide Field Camera 3 medium- and narrow-band filters}

In addition to the medium and wide filters commonly used in extragalactic surveys, there are some medium- and narrow-band filters to consider. These are mostly within the F125W and F140W wide-band filters:  F126N, F127M, F128N, F130N, F132N, F139M, and F153M. Some of these will be centered on the broad absorption features in red or brown dwarf spectra. A similar approach using a wide and a narrow filter has worked well to determine the low-mass end of the stellar initial mass function in Galactic star-formation regions \citep{Najita00, Andersen06, Da-Rio12a}. Narrow-band imaging filters 
(e.g., J1 or J2 medium-band filters) centered on the methane bands can well be used to type the lower-mass dwarfs \citep{Tinney12}.

\subsection{\emph{James Webb Space Telescope}}

The NIRCam instrument on board the \emph{JWST} is a versatile instrument with a range of broad, medium and narrow-band filters. We evaluate the NIRCam F070W, F090W, F115W, F140M, F150W, F150W2, F162M, and F200W filters (Figure \ref{f:filters}). NIRCam also has longer wavelength filters, but we cannot evaluate those filters because of a lack of coverage by the spectra in {\sc SPLAT}. In Holwerda et al. ({\em in prep}), we found that Spitzer [3.6] - [4.5] $\mu$m color correlates well with spectral type for the cooler ($>$ L5) brown dwarfs \citep[see also][]{Kirkpatrick11, Pecaut13, Skrzypek16}. It is therefore possible that the longer wavelength filters on NIRCam and the MIRI instrument hold possibilities for improved brown dwarf characterization.

\subsection{\emph{Euclid}}

The European Space Agency \emph{Euclid} mission has three main goals: it will be used as a gravitational lensing experiment, a photometric redshift experiment to map baryon acoustic oscillations, and a supernovae search \citep{EUCLID}. 
The design calls for only three filters ($Y$, $J$, and $H$; Figure \ref{f:filters}) in the near-infrared, and a very wide-band filter in the optical (for the gravitational lensing measurements). 

For the purposes of this work, the gain from this observatory is that it will
scan the entire sky to a depth of $m_J=24$ and the ecliptic poles to a depth of $m_J=26$, making \emph{Euclid} photometry universally available in combination with any other observatory's filters. The NIR filters for \emph{Euclid} are wide Y, J and H filters modeled with a simple wavelength window. The PSF is $0.2$" in the NIR filters.

\subsection{\emph{WFIRST}}

\emph{WFIRST} \citep{Dressler12,WFIRST,Thompson13a} is NASA's next flagship astrophysical mission. It will have a limited selection of broad near-infrared filters: F184, H158, J129, W146, Y106, Z087, and F062 for the Wide Field Instrument (Figure \ref{f:filters}). The WFI provides an imaging mode covering 0.76 -- 2.0 $\mu$m and a spectroscopy mode covering 1.35 -- 1.95 $\mu$m. The wide field focal plane covers an effective field of view of 0.281 deg$^2$ with a 0.11" pixel scale and 0.2" PSF width.

\subsection{Morphological identification of stars}

All four observatories sample the width of their PSF by approximately two pixels. Our experience with \emph{HST} suggests that this is enough for a reliable morphological identification of stars $\sim$1.5 magnitudes above the photometric limit  of a given survey using the effective radius determined from a growth curve \citep[the ranked list of pixels in each object; see ][]{Ryan11,Holwerda14}. Better sampling improves the spread in effective radius for stars into a tighter relation with luminosity. However, the limiting magnitude for differentiating between stars and galaxies is only extended by $\sim 0.2$ magnitude. Given that our target is the substellar population of Milky Way dwarfs, the morphological selection is close to identical for different instruments. 

\subsection{Grism surveys}

The scope of this paper is limited to imaging surveys. Specifically, we evaluate whether or not the combination of existing and planned surveys can readily identify red or brown dwarf populations, or how a small filter addition can convert such surveys into an optimal dwarf census. However, these four observatories all have grism low-resolution spectroscopy capabilities. 

\emph{HST}/WFC3 and ACS grism observations have already been used to identify brown/red dwarf stars in our Galaxy; \cite{Pirzkal05,Pirzkal09} reported M-dwarfs in ACS grism data, and \cite{Masters12a} reported the discovery of three late $>$ T4.5-type dwarfs in the WFC3 Infrared Spectroscopic Parallels (WISP) survey. Similarly, one can expect more substellar objects to be subtyped in the 3D-HST \citep{Brammer12, Skelton14, Momcheva16a} and FIGS \citep{Tilvi16, Pirzkal17} WFC3 grism surveys. The limiting magnitude on shallow grism observations is effectively $m_J=23$ \citep{Masters12a} and well over two magnitudes deeper in similar time investment in pure-parallel imaging observations \citep{Ryan11,Holwerda14}. 

\emph{WFIRST} will observe $\sim$ 2000 deg$^2$ with its grism element to depths of $m_{AB}\sim 26$ with a wavelength range of 1.35 -- 1.89 $\mu$m \citep{Spergel13,Spergel15}. In effect this will be an ideal brown/red dwarf data-set, covering at least one of the wide absorption features that define the later brown dwarf types (Figure \ref{f:standards}). 

In this paper, we do not consider these grism observations only as matter of strategy and a way to limit the scope of the work. Our considerations for not including grism comparisons are the following: (1) imaging observations are easier to compare across instruments for their efficiency at typing red or brown dwarfs, (2) the aim of the paper is to ascertain whether certain extragalactic programs can be used for red or brown dwarf typing as is or require an amended observing strategy, and (3) the possible use of future pure-parallel observations similar to BoRG (by \emph{HST}, \emph{JWST} or \emph{WFIRST}) in the typing of brown/red dwarfs and the scale of the Milky Way.

\subsection{Proper motion identification}

Many of the red and brown dwarfs in the solar neighborhood have been identified using their proper motion \citep{Kirkpatrick14, Kirkpatrick16,Robert16, Kuchner17}. In the case of the higher latitude imaging surveys, this may still be possible but it will be proportionally more difficult because proper motions for these more distant, fainter objects are expected to be on the order of milliarcseconds, especially in the case of halo objects.
\cite{Ryan05} used proper motion to help identify M-type dwarfs and this could be a viable verification method, depending on the observing strategy of the extragalactic surveys (ideally multiple, well-spaced epochs). In reality, cadence and PSF-centering are dictated by spacecraft limits and searches for extragalactic transients (e.g., supernovae), and they may be well-suited for proper motion studies. 

Imaging surveys are more efficient at covering larger areas to a greater depth and with multiple pointings than grism surveys, although they sacrifice some accuracy in dwarf typing in the process. To map the structure of the Galaxy in dwarf stars, multiple lines-of-sight and a large volume will be necessary.

\section{Brown dwarf selection}
\label{s:2col}

\cite{Ryan11} and \cite{Holwerda14} used two near-infrared colors observed with WFC3 to identify objects as dwarf stars, separating them from high-redshift objects. Subsequently, these near-infrared colors can be used to identify the broad brown dwarf types. By necessity, these were the F098M, F125W and F160W of the BoRG survey \citep{Trenti11}. Stellar objects that were already morphologically identified were typed further as M-type dwarfs, and then subtyped using an optical-near-infrared color.  
Here, we first evaluate different two color filter combinations to separate out broad dwarf star type. To evaluate the resolving power of a single color mapping to red or brown type, we use the Spearmann ranking ($\rho$). If type increases or decreases monotonically with color, the Spearmann ranking is close to unity ($\pm1$), representing an ideal case to use that color for subtyping.

\begin{figure*}
\includegraphics[width=0.49\textwidth]{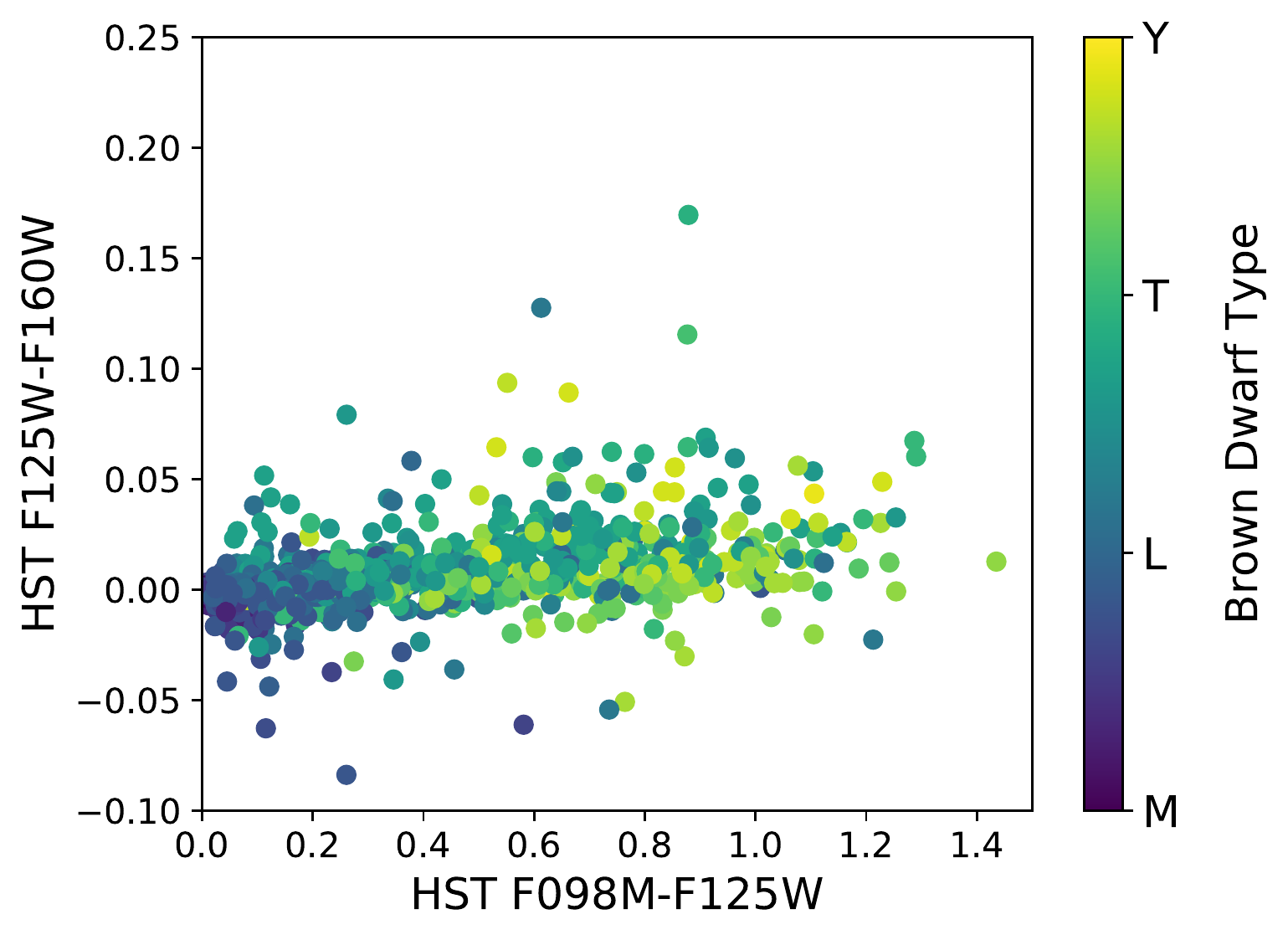}
\includegraphics[width=0.49\textwidth]{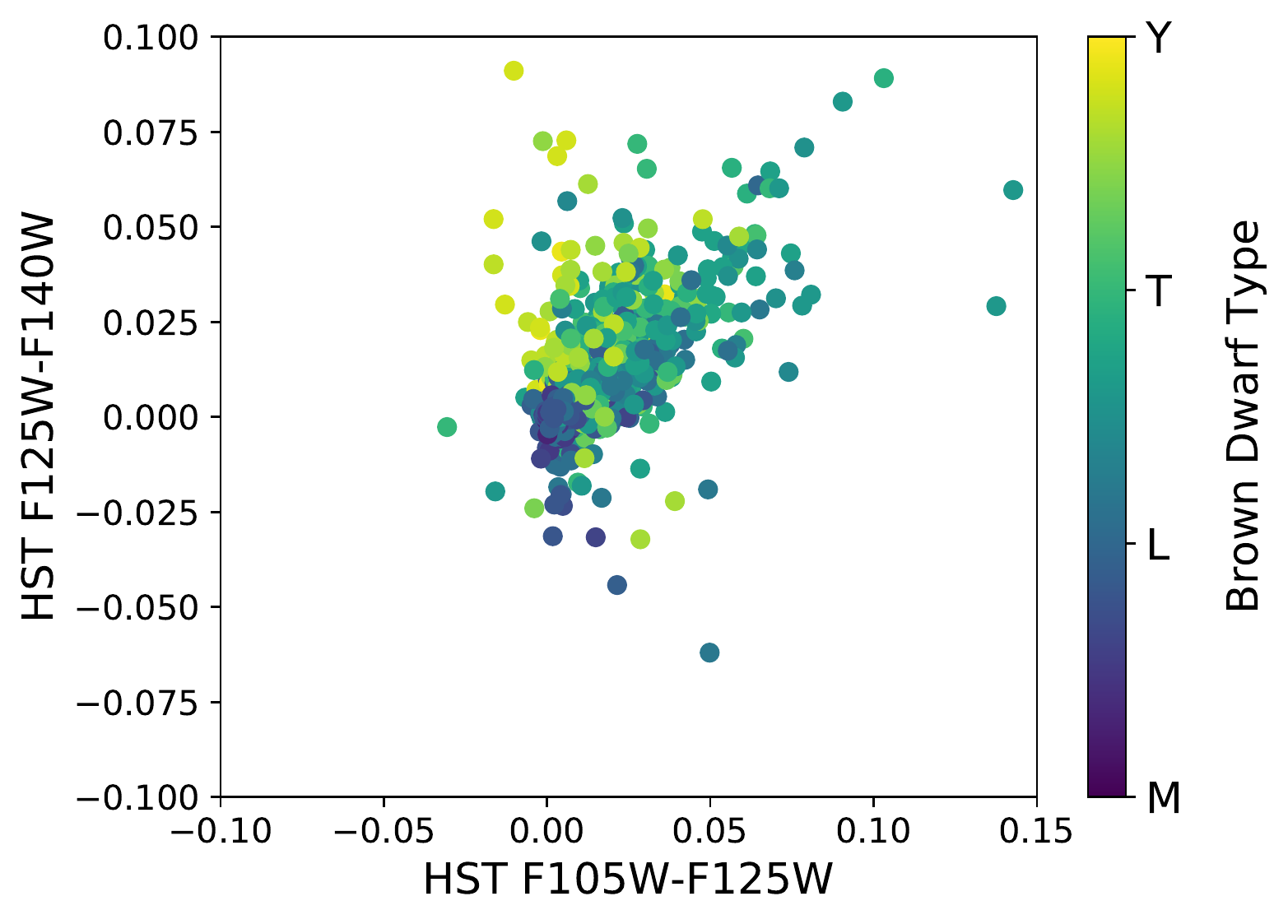}
\caption{\label{f:hst:2col}\emph{HST} color-color combinations for brown dwarf selection as a class of objects. These are is the common filter combinations for \emph{HST} pure-parallel observations \protect\citep{Ryan11, Holwerda14,van-Vledder16}. }
\end{figure*}


Table \ref{t:2col} shows the Spearman ranking between the brown dwarf type and \emph{HST} color pairs. The best combinations are $Y_{F105W}$--$J_{F125W}$ combined with $J_{F125W}$--$JH_{F140M}$, when only considering broad filters, often employed for extra-galactic work. This is remarkably close to the BoRG[z9] filter combinations \citep{Calvi16,Bernard16} but differs from the $Y_{F098M}$--$J_{F125W}$/$J_{F125W}$--$H_{F160W}$ filter combinations used earlier in \emph{Hubble} pure-parallel observations \citep{Holwerda14,van-Vledder16}. The difference in Spearman ranking between these two filter combinations is only moderate (Table \ref{t:2col}); both combinations perform similarly in distinguishing broad types.

\begin{figure}
\includegraphics[width=0.5\textwidth]{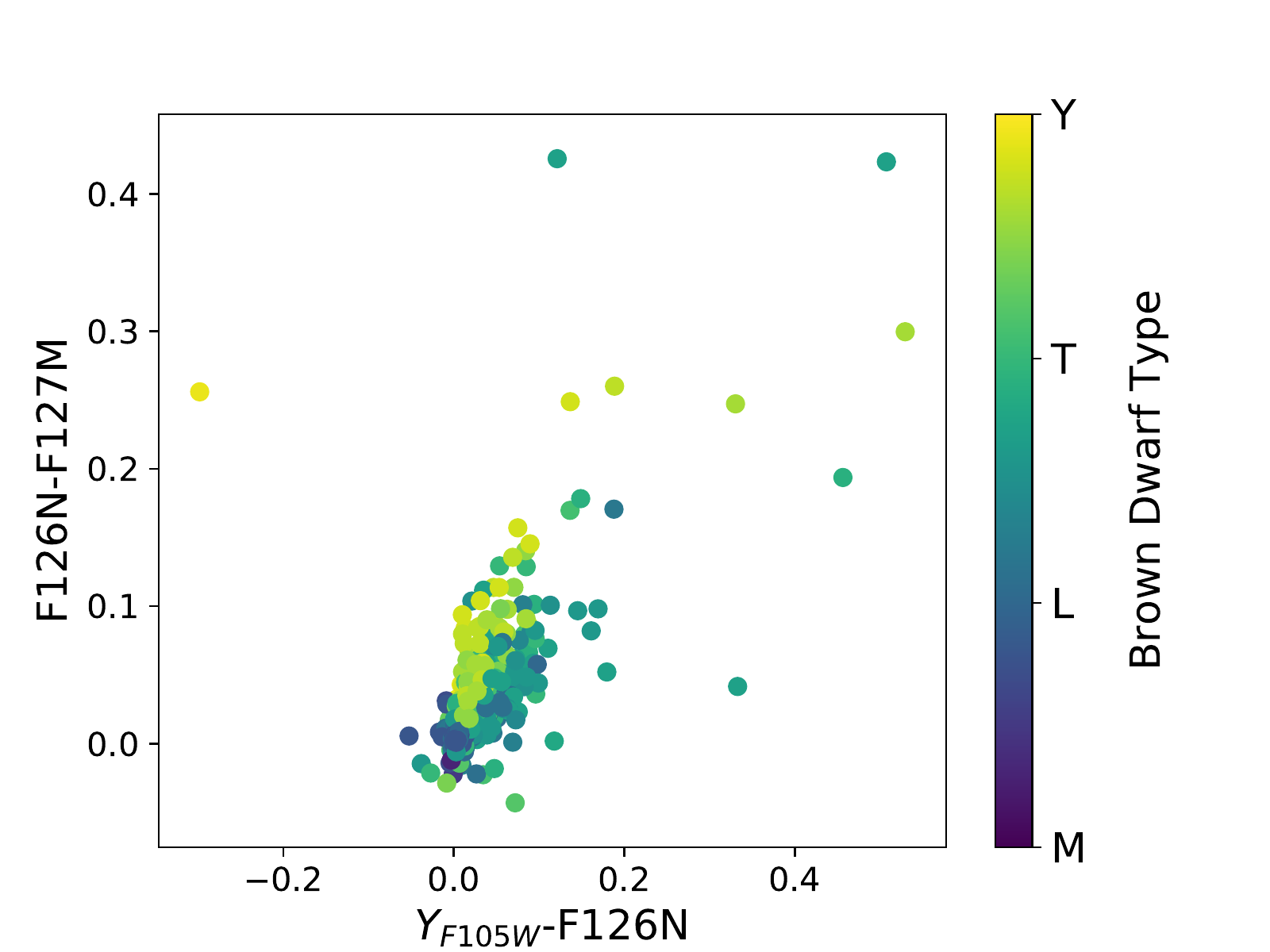}
\caption{\label{f:hst2:2col:med} Optimal \emph{HST} color-color selection including medium and narrow-band filters not often used in extragalactic surveys. Separation into broad categories is not dissimilar to the performance of the broad-band filters already in use.}
\end{figure}

Figure \ref{f:hst:2col} shows the optimal HST filter combinations identified by the Spearman ranking for separating out dwarf types as well as the color-color combination used most often so far, i.e. based on existing data. Both discriminate the broad brown dwarf populations reasonably well. Ongoing searches for brown dwarfs in BoRG WFC3 fields will be able to reliably type the brown dwarfs using either filter combination.

We can expand the color possibilities if we include the medium- and narrow-band filters that are not typically used in extragalactic surveys. The two-color selection improves somewhat when one medium filter often used in extragalactic surveys (F105W) is combined with two filters not commonly used in extragalactic filters (F126N, F127M; Figure \ref{f:hst2:2col:med}). Separation of the dwarf broad types is somewhat better with this filter combination than the extragalactic filter combinations (Figure \ref{f:hst:2col} and \ref{f:hst2:2col:med}). We note, however, that once this is applied to the full {\sc SPLAT} sample of red or brown dwarfs, the relation is not as clear cut.

\begin{figure}
\includegraphics[width=0.5\textwidth]{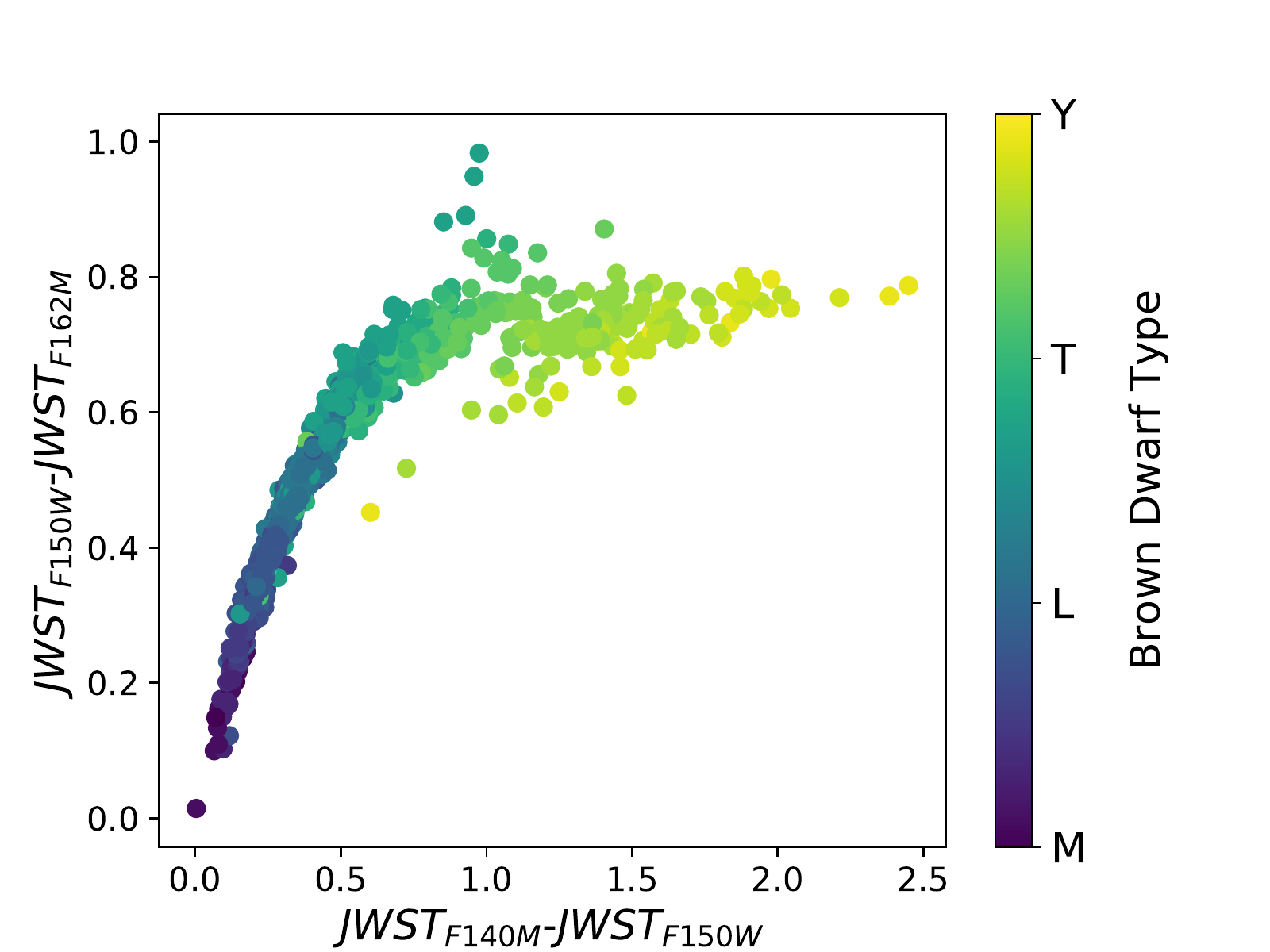}
\caption{\label{f:jwst:2col}Relation between \emph{JWST} F140W, F150W, and F162M colors and the brown dwarf type and subtype for all the objects in the {\sc splat} catalog. The combined broad/medium filter set separates out the type and subtype very well for all, outperforming the HST F125W/F127M filter combination.
}
\end{figure}

\begin{figure} 
\includegraphics[width=0.5\textwidth]{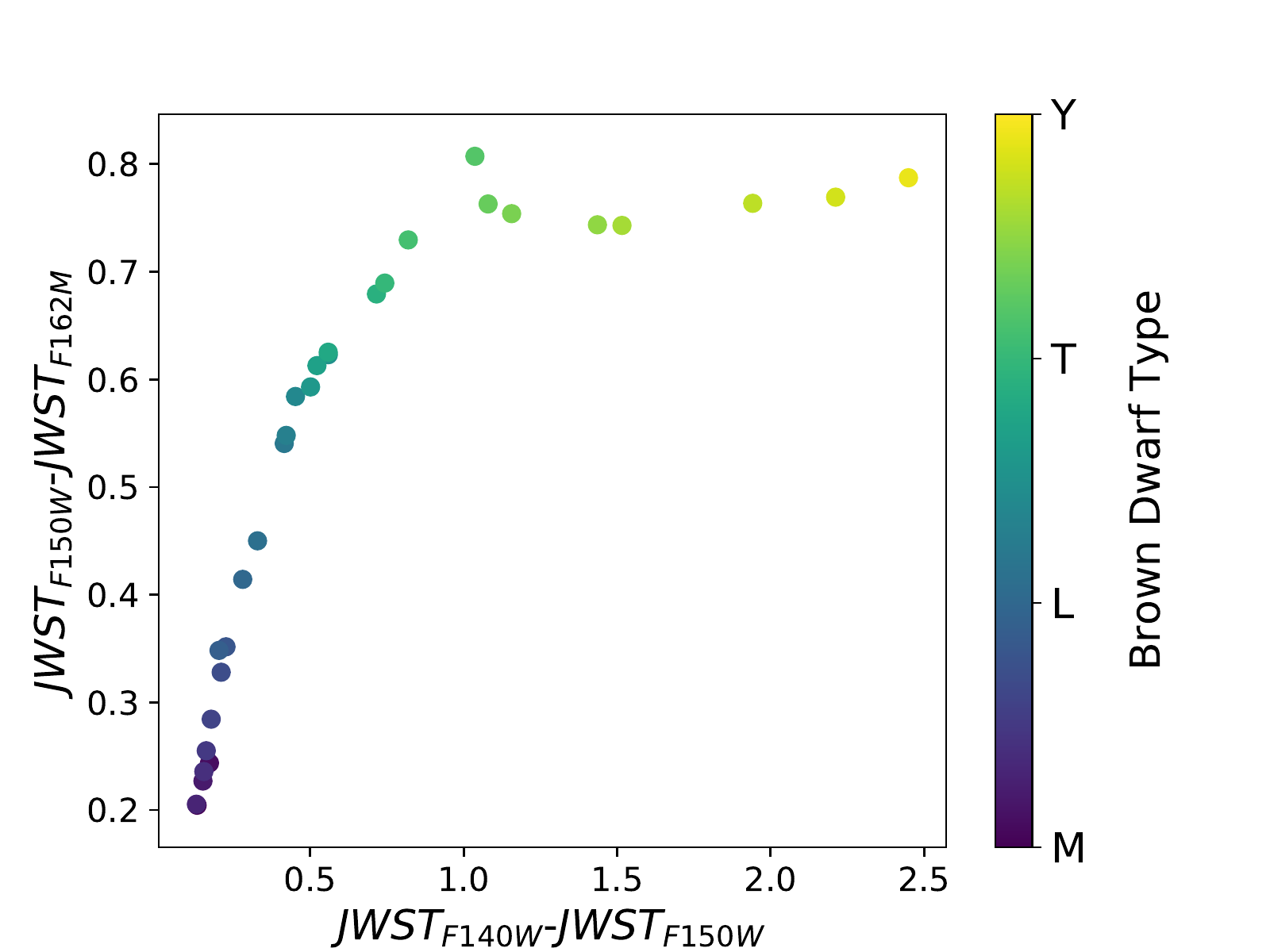}
\caption{\label{f:jwst:std:2col}\emph{JWST} color-color selections for brown dwarf standard stars. }
\end{figure}

The many filters available on \emph{JWST} offer the opportunity to discriminate among brown dwarf types and subtypes, as they span the deep absorption features in the later types. The best  combination to  discern between  brown dwarf types is  $F140M$--$F150W$ and $F150W$--$F162M$ (Figure \ref{f:jwst:2col}). This filter combination cleanly separates the different dwarf subtypes with a clear progression from early to late (M0 to T8) along these two colors (Figure \ref{f:jwst:std:2col}). This strongly illustrates the need for a medium band filter to subtype brown dwarfs and the power of a well-placed medium band filter to do so.

Figure \ref{f:jwst:std:2col} shows the same plot but for only the {\sc SPLAT} standard stars (i.e., those that define the type) from \cite{Burgasser06}, \cite{Kirkpatrick10}, and \protect\cite{Cushing11}. This figure demonstrates how the separation in color space allows subtyping without significant degeneracy.

\begin{figure} 
\includegraphics[width=0.5\textwidth]{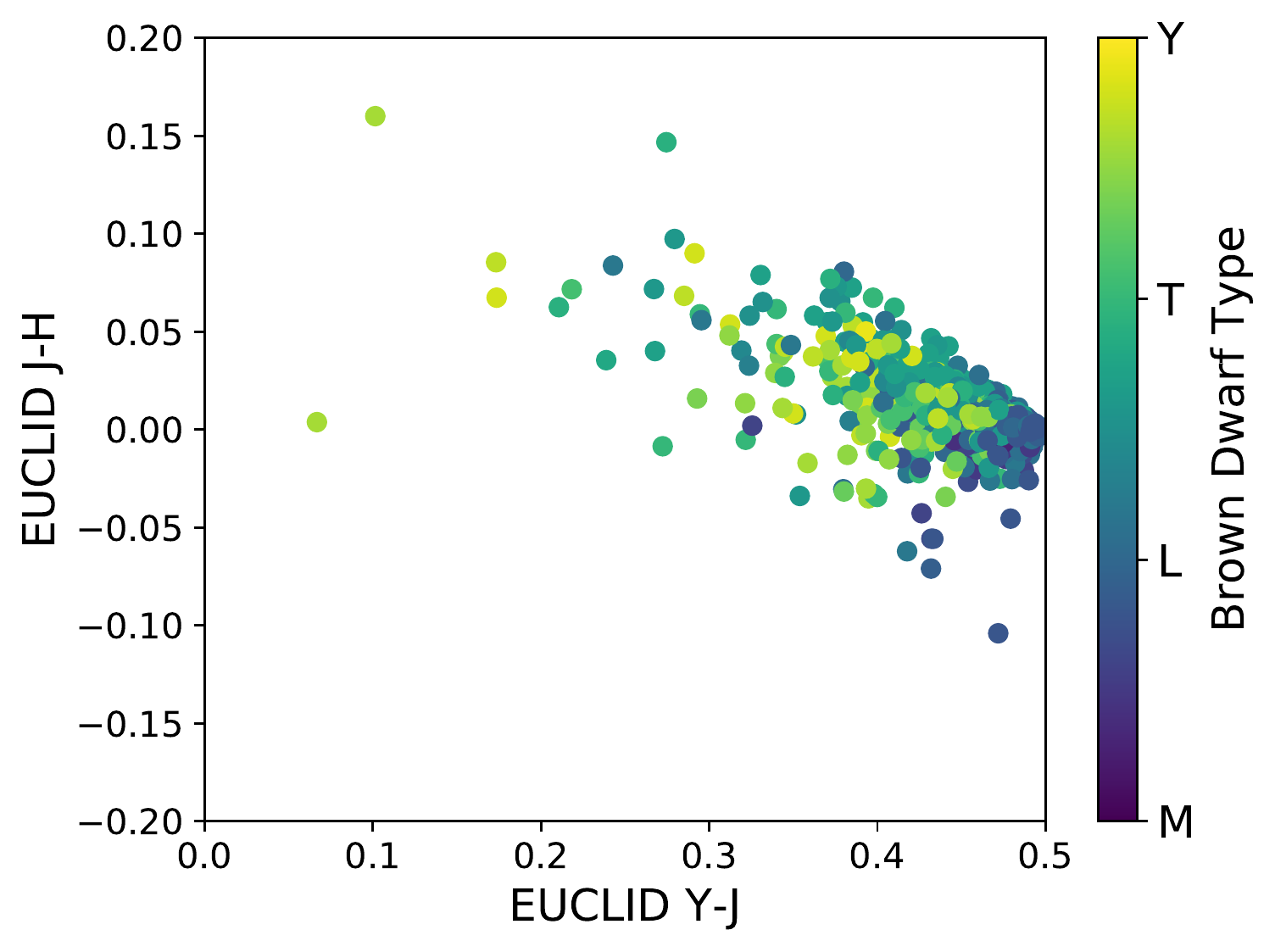}
\caption{\label{f:euclid:2col}The \emph{Euclid} color-color plot of all the brown dwarfs in {\sc SPLAT}. The broad filters of \emph{Euclid} identify brown dwarfs as a population, but do not discriminate between broad classes.}
\end{figure}

\begin{figure}
\includegraphics[width=0.5\textwidth]{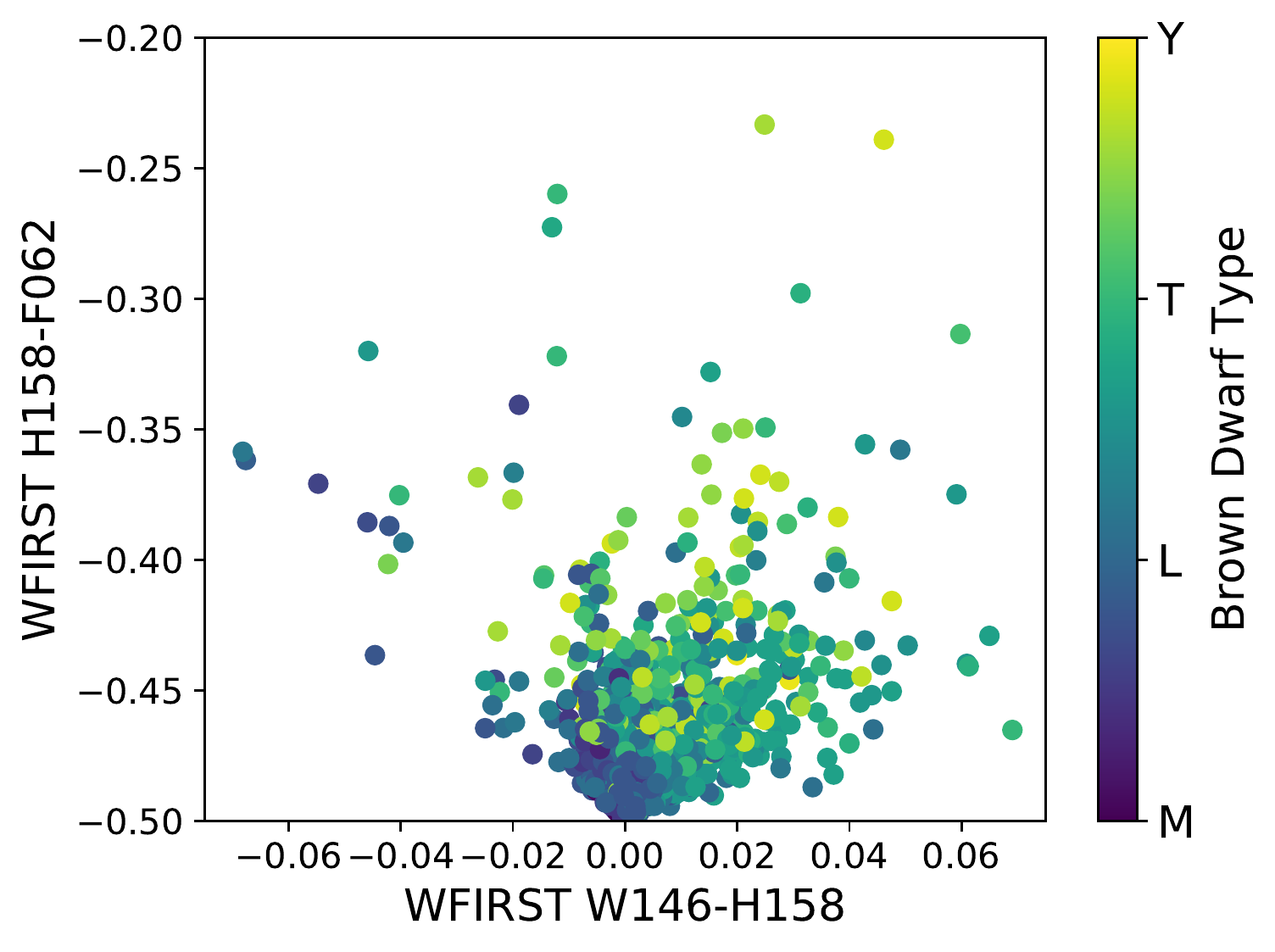}
\caption{\label{f:wfirst:2col}\emph{WFIRST} color-color plot of all the brown dwarfs in {\sc SPLAT}. The W146, H158 and F062 filters separate out  the broad brown dwarf types --M, L, or T? -- reasonably well, similar to the \emph{HST} broad band filters.}
\end{figure}

The broad \emph{Euclid} filter set identifies a clear window in color-color space to identify red or brown dwarfs as a class of objects. However, these bands do not discriminate between broad type (Figure \ref{f:euclid:2col}). 
The near all-sky coverage of this mission will therefore result in an ideal parent sample to explore the Milky Way structure; dwarf stars will be relatively easily identified as a class of objects. However, to identify in detail what type of dwarf star each star is, it will require follow-up observations such as \emph{Euclid} grism sub-type identification or a proper motion measurement between \emph{Euclid} observation epochs to determine distance and hence likely type and subtype. 

Like \emph{Euclid}, \emph{WFIRST}/WFI is a wide-area survey instrument, designed to observe many high-latitude fields. Unlike \emph{Euclid}, there are more filters available to discriminate among brown dwarf types. They do so with intermediate success, similar to the broad \emph{HST} filters used so far. 
There are weak trends with broad dwarf type in many filter combinations (see Table \ref{t:2col} and Figure \ref{f:wfirst:2col}).  Most of the brown dwarfs are tightly grouped in color-color space (i.e., a narrow range of values in W146-H158, see Figure \ref{f:wfirst:2col}) with a few  outliers.


\begin{figure}
\includegraphics[width=0.5\textwidth]{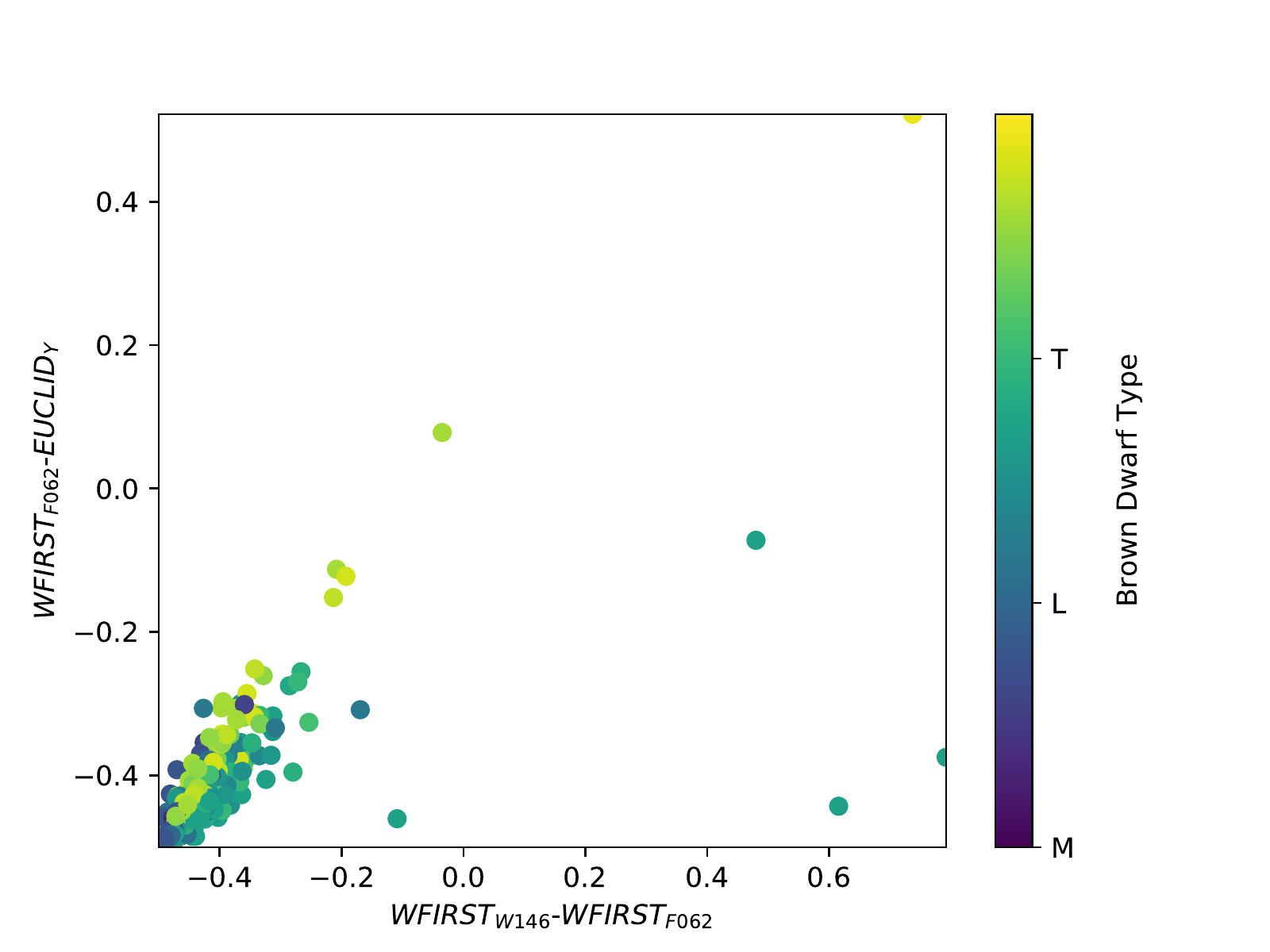}
\caption{\label{f:euclid:wfirst:2col}\emph{Euclid} and \emph{WFIRST} color-color plot of all the brown dwarfs in {\sc SPLAT}. The W146 and F062 filters on \emph{WFIRST} combined with the \emph{Euclid} $Y$-band distinguish reasonably well the broad brown dwarf types with some conspicuous outliers.}
\end{figure}

The mission concepts for both \emph{Euclid} and \emph{WFIRST} are in an advanced stage, including choices for the filter combinations. It would therefore be fortuitous if a combination of \emph{WFIRST} and \emph{Euclid} filters could perform well in the subtyping of substellar dwarf. \emph{Euclid} will cover the entire extragalactic sky and the \emph{WFIRST} science case includes a very wide survey. Photometry by both mission is nearly guaranteed. 
However, the performance of \emph{Euclid}/\emph{WFIRST} filter combinations is a qualified success: \emph{WFIRST}'s W146 and F062 in combination with \emph{Euclid}'s $Y$-band filter performs a better separation of general type than either observatory individually (see Figure \ref{f:euclid:wfirst:2col}).

\section{Dwarf subtyping}
\label{s:1col}

Brown dwarf subtyping may be achievable using a single color that has not yet been used to identify the broad type. The idea is that a combination of three (broad) filters is used to subdivide stellar objects into M/L/T-type objects and one additional filter allows for the discrimination between, for example M3 and M5.
For instance, \cite{Holwerda14} used one separate color (V-J) to subtype M-type dwarfs in extragalactic fields. However, \cite{van-Vledder16} note that this kind of subtyping lets in a high level of contamination ($\sim$50\%) from other subtypes, enough to smooth out scale height differences between subtypes.

\begin{figure}
\includegraphics[width=0.5\textwidth]{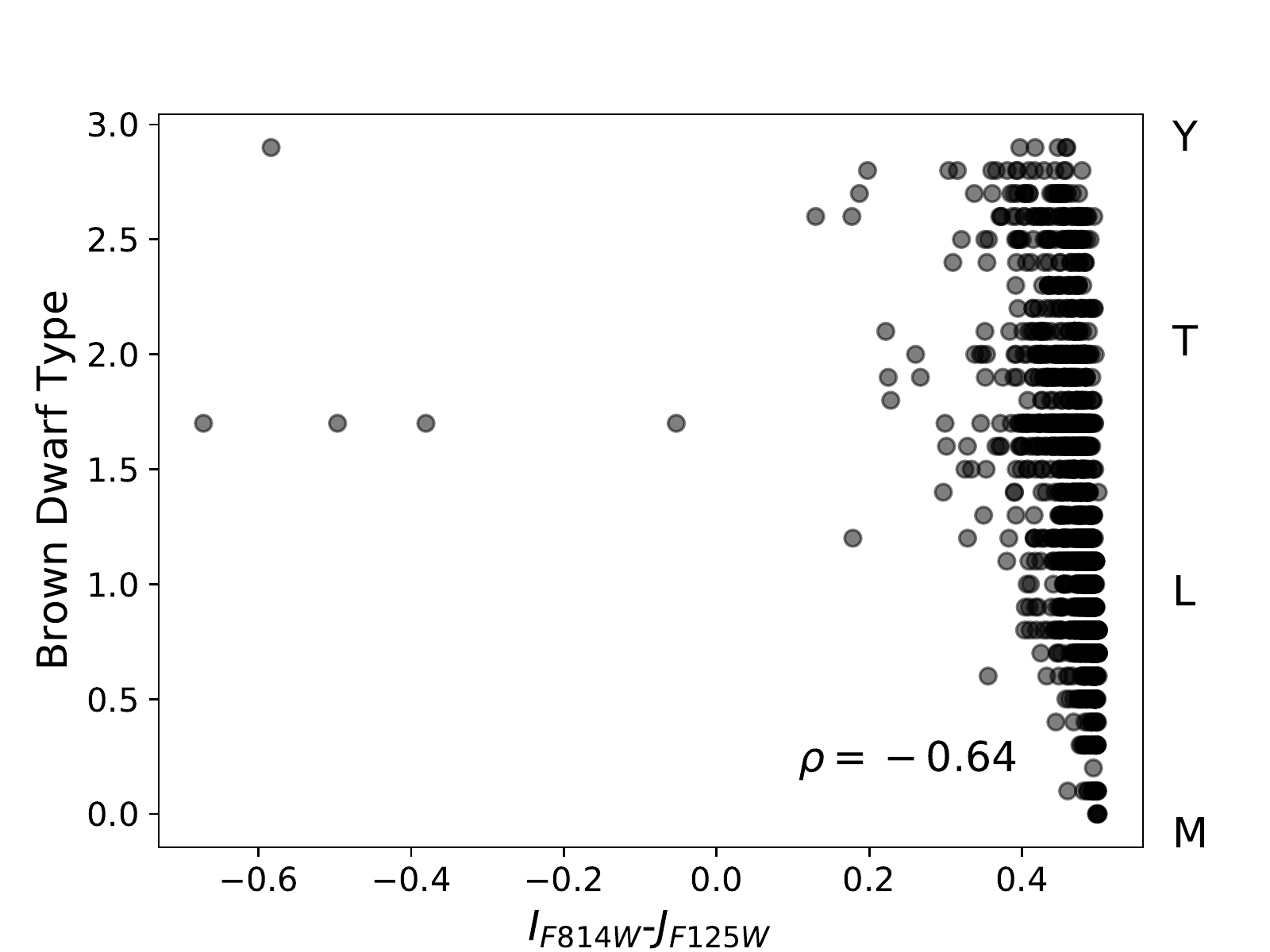}
\caption{\label{f:hst:1col}\emph{HST} color-type relation for brown dwarfs. red or brown dwarf types are numerical as follows: M0-M10 (0-1), L0-10 (2-3), T0-T10 (2-3) and Y0-Y3 (3-3.3). \protect\cite{Holwerda14} used a optical-near-infrared color (V-J) to subtype M-dwarfs. A optical-near-infrared color using broad filters may work well enough for M-dwarfs as type correlates with black body temperature but later types become degenerate in color space (e.g., L and T), the broad absorption features counteract the temperature effect. This makes (sub)typing difficult with only broad filters. }
\end{figure}

In the case of local red or brown dwarfs, the near-infrared color (e.g., the WISE W1-W2 or Spitzer [3.6]-[4.5] $\mu$m colors) can be used to probe the Rayleigh-Jeans part of the dwarf spectra and hence type them accurately photometrically \citep[e.g.,][]{Kirkpatrick11, Pecaut13, Skrzypek16}. Similarly, \cite{Labbe06} used Spitzer colors to reject dwarfs from their extragalactic searches. Barring deep Spitzer photometry or \emph{JWST}/NIRcam observations in the 3-5 $\mu$m range, this information is not available for deep extragalactic imaging campaigns. We focus here on near-infrared filter combinations that relate linearly to type, avoiding color degeneracies.

Figure \ref{f:hst:1col} shows an example of a color-type relation. To quantify the relationship between a near-infrared color and brown dwarf (sub)type, we use the Spearman ranking of the values. If there is a monotonically increasing or decreasing relationship, the Spearman ranking ($\rho$) is close to 1 or -1 respectively. If no such relation exists, the value is close to 0. We report Spearman rankings for the full {\sc SPLAT} sample and the standard stars.

\begin{figure}
\includegraphics[width=0.5\textwidth]{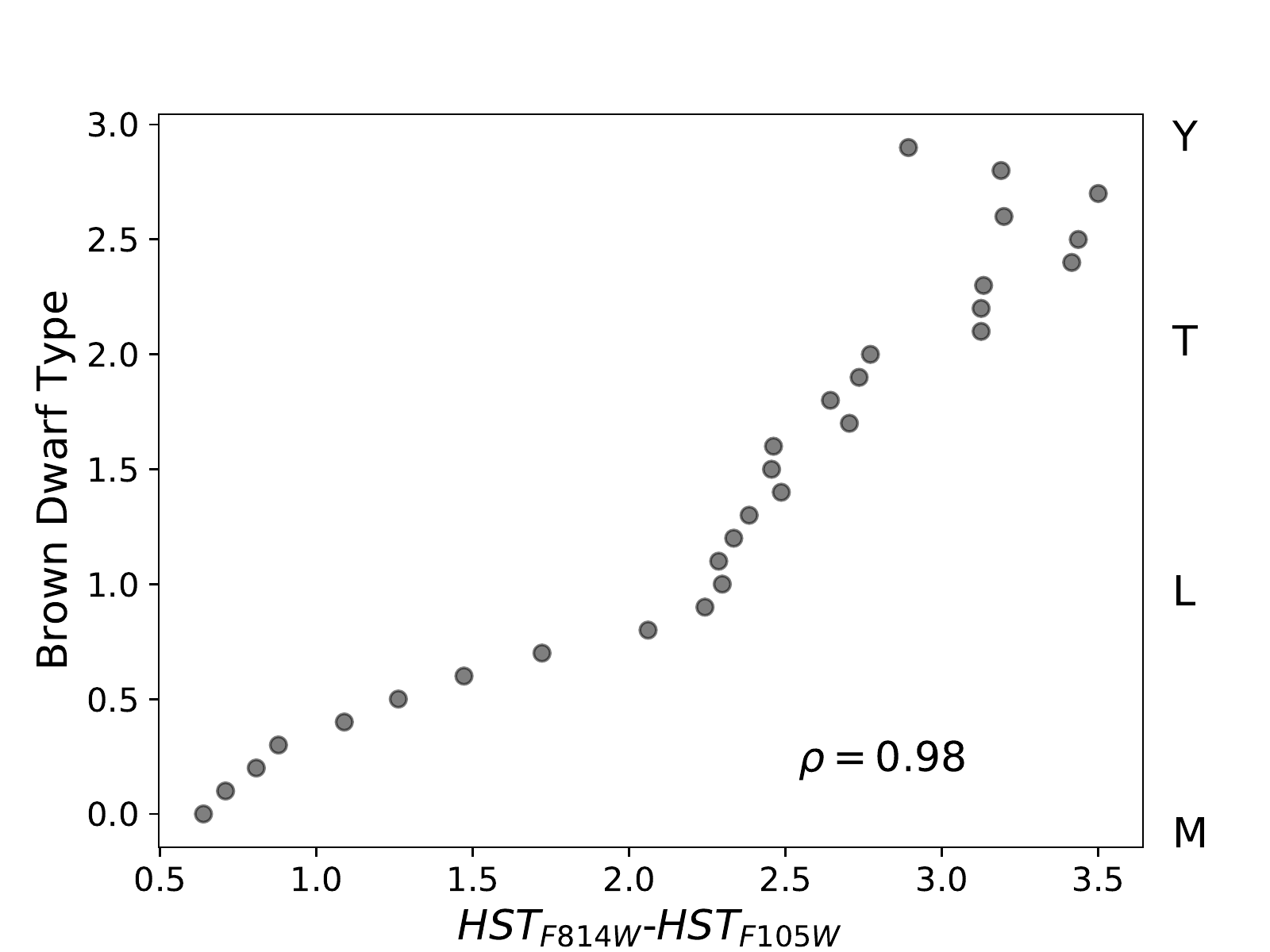}
\caption{\label{f:HST:YI}F814W--F105W color as a function of type for the standard red or brown dwarfs in {\sc SPLAT}. Types marked similar to Figure \ref{f:hst:1col}. This optical-near-infrared broad filter combination comes closest to subtyping dwarf and substellar objects but is mostly not degenerate for M-dwarfs.}
\end{figure}

In the case of \emph{HST} imaging, \cite{Holwerda14} used the $V$-$J$ color (F606W-F125W) to subtype the M-dwarfs once they were selected using the F098M-F125W/F125W-F160W color space. Despite some success with sub-typing, this relationship needs to be recalibrated with every new survey's filter choice targeting higher redshift objects as their choice of optical filter (to verify the dropout) continued to change. 
Of all the broad filters commonly used in extragalactic surveys that were evaluated, the F184W--F105W combination ($\rho=0.98$) has the best correlation with subtype, but similar to the $V$-$J$ color used in \cite{Holwerda14}, the optical-near-infrared color works well for M-dwarfs but becomes degenerate for later types (L+, see Figure \ref{f:HST:YI} for the relation with standard stars).

However, we now consider the medium- and narrow-band filters not commonly used in extragalactic surveys. Many combinations between a wide-band filter and one of these narrow-band filters result in a very strong correlation with red or brown dwarf spectral type. Table 
\ref{t:1col} lists the wide-/narrow-band filter combinations one could consider. The best-performing filter combination is F105W-F126N. However, the F105W filter is not always used in typical extragalactic surveys such as BoRG and CANDELS. F125W and F160W are the most commonly available. For F125W ($J$), either F127M would work almost as well as the optimal F105W-F126N combination. For F160W ($H$), the optimal narrow-band filter pairing is F153M. 
\begin{figure*}
\includegraphics[width=0.49\textwidth]{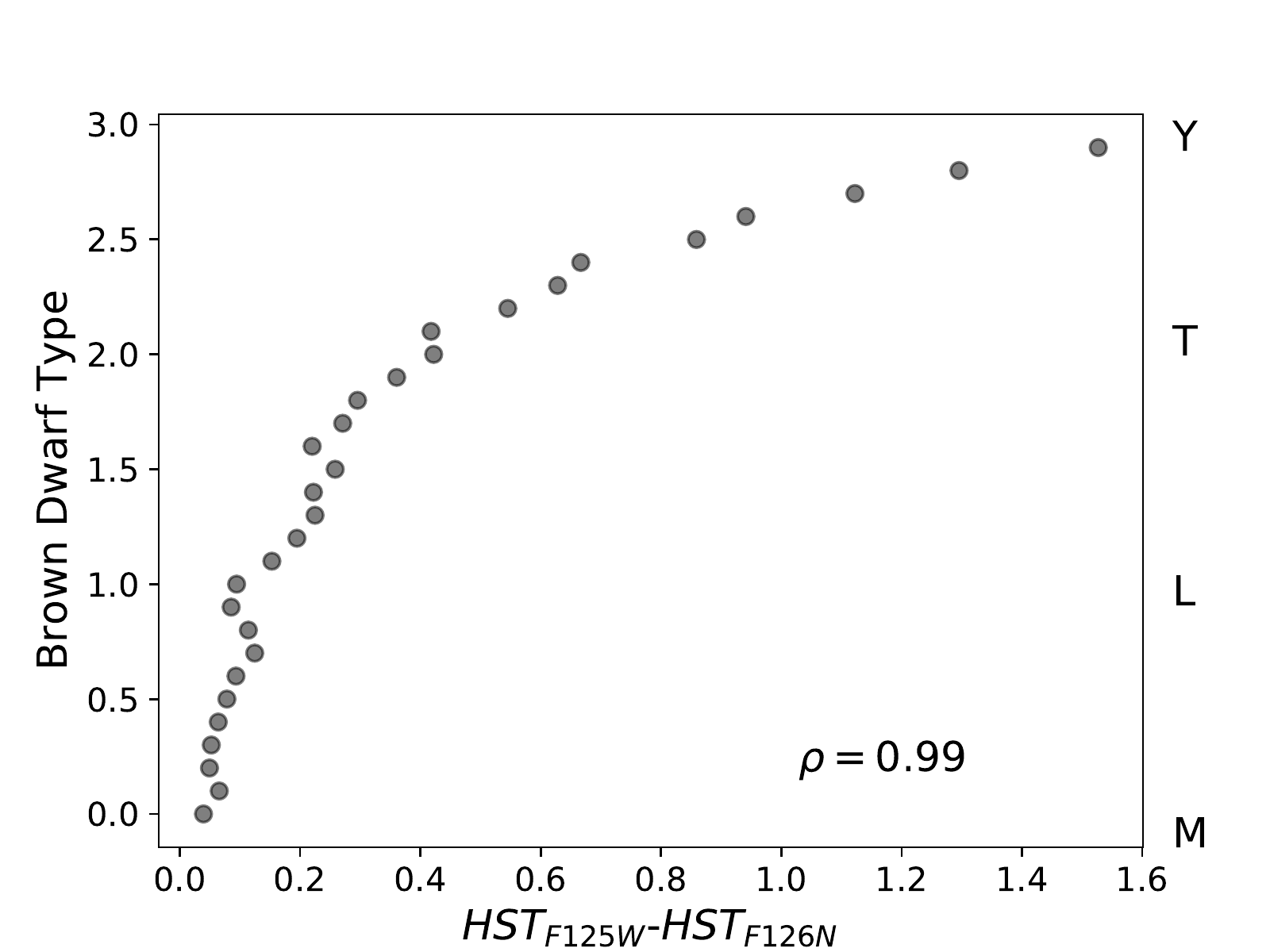}
\includegraphics[width=0.49\textwidth]{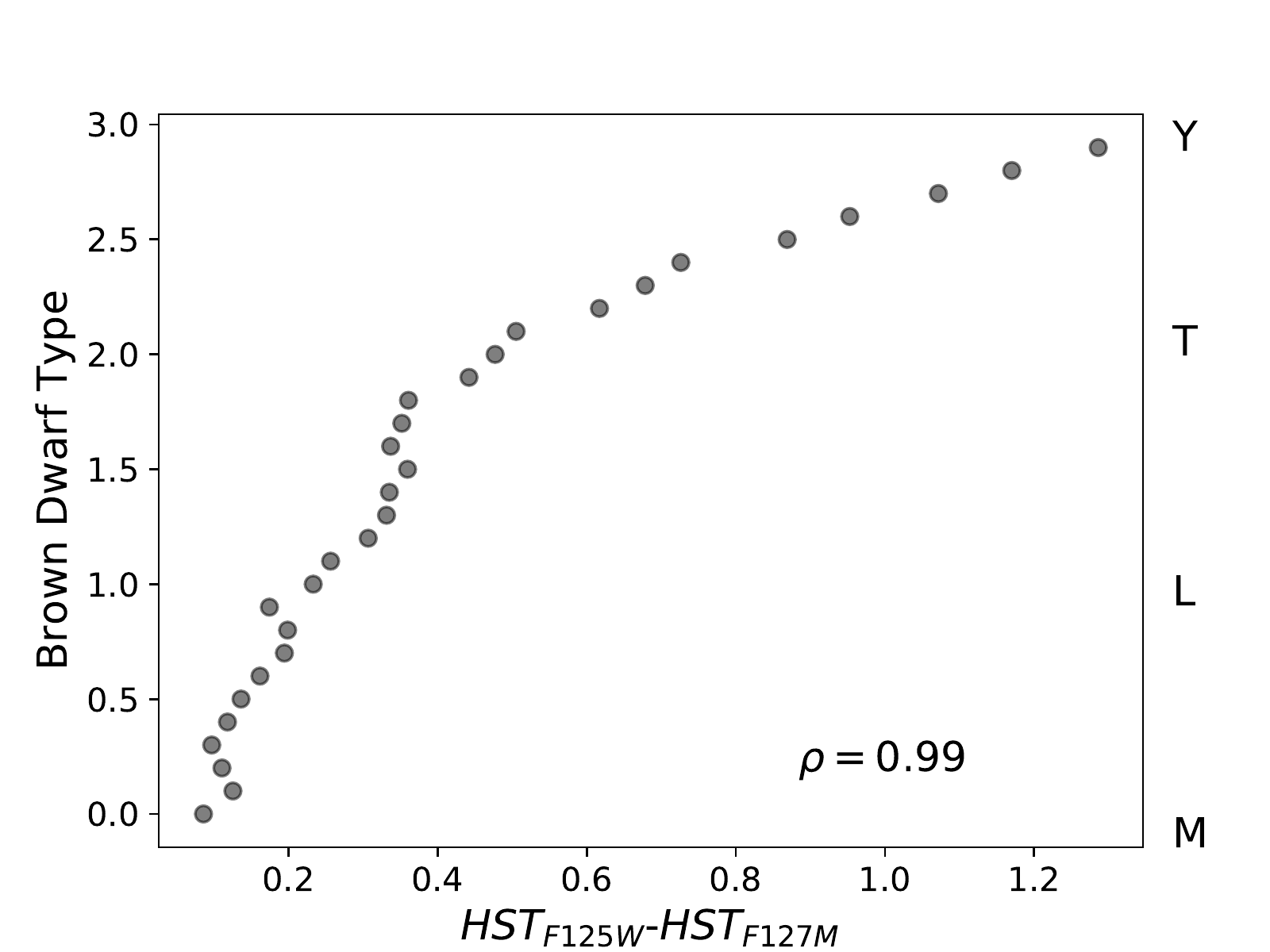}
\includegraphics[width=0.49\textwidth]{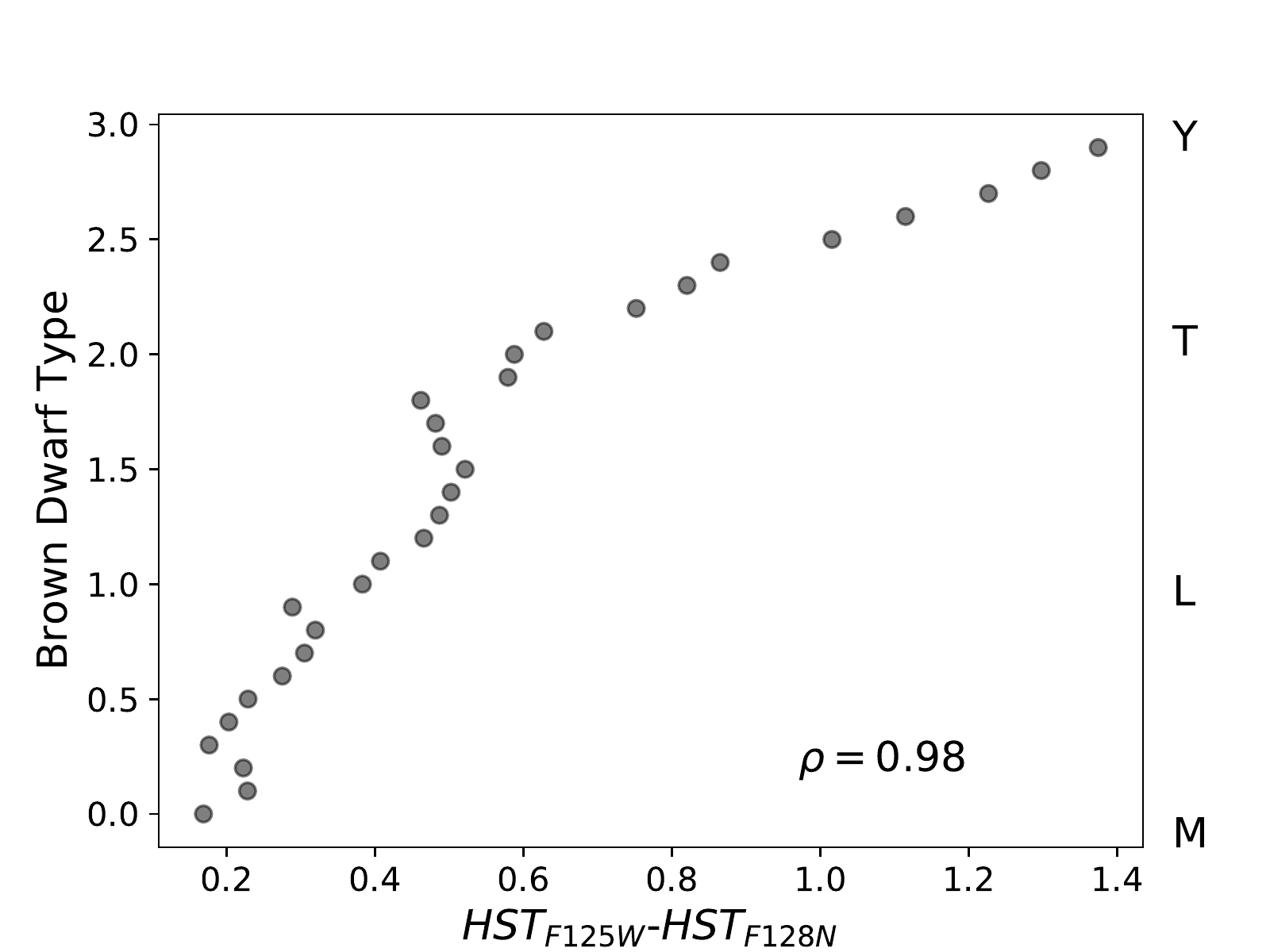}
\includegraphics[width=0.49\textwidth]{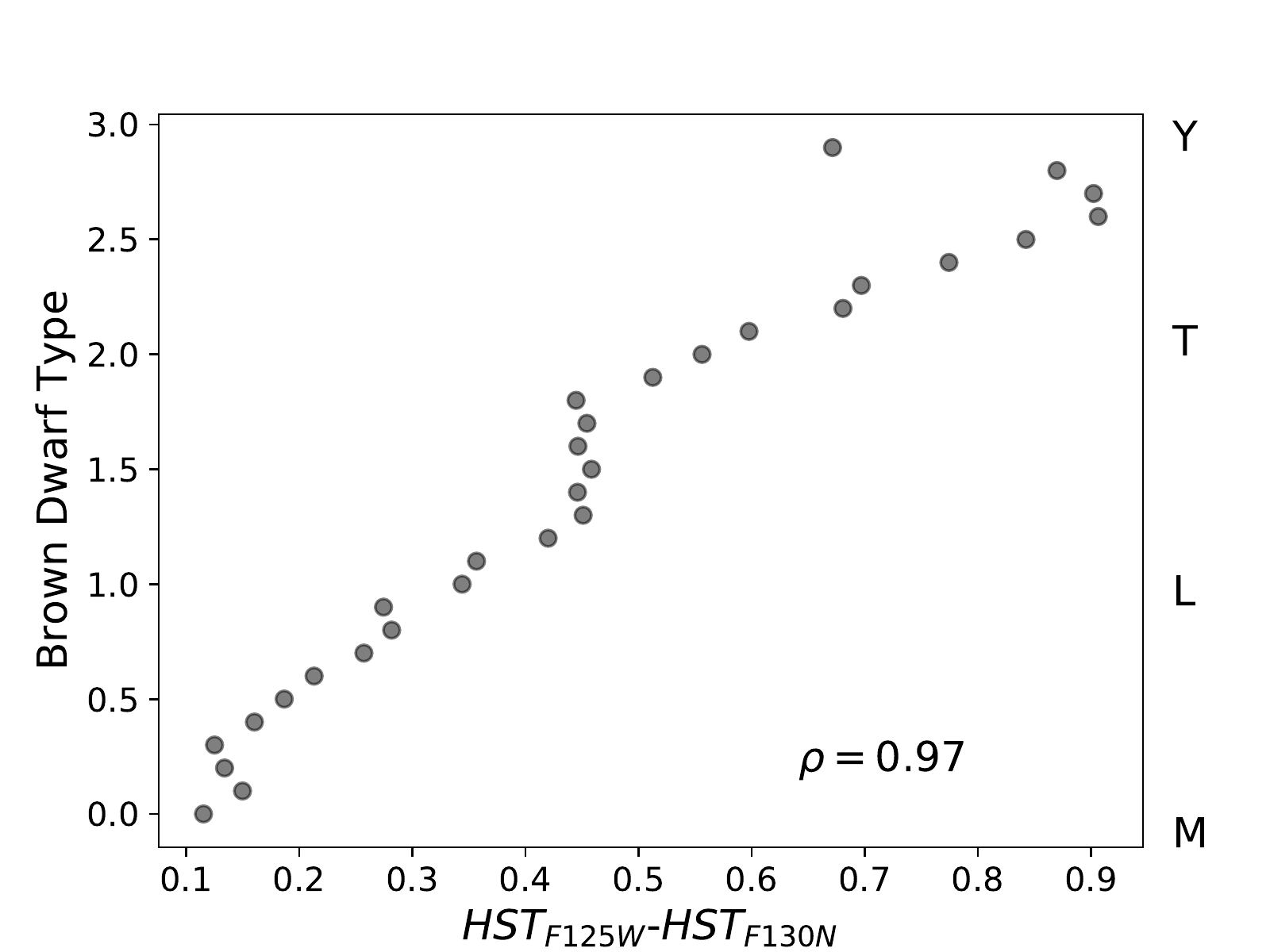}
\caption{\label{f:hst:nb2:1col} Four plots showing the discriminatory power of the combination of an HST broad-band filter (F125W) with a narrow or medium band one for the standard stars in {\sc splat}: F126N (top left panel), F127M (top right panel), F128N (bottom left panel) and F130N (bottom right panel). Most extra-galactic observations feature the F125W. This is to show which medium or narrow band filter should be added to that to improve the (sub)typing of the brown dwarfs in these fields. All work well ($\rho>$0.90) but the F127M stand out as an excellent compromise with a monotonous relation between the F125w-F127M color and type and highest throughput of all the considered narrow/medium bands.}
\end{figure*}

To turn existing extragalactic surveys into a fast red or brown dwarf surveys of the Milky Way disk by including a single medium or narrow-band filter not previously considered, one should consider including either F126N or F127M as either one would result in excellent subtyping. 
Apart from the Spearman rankings, the F125W--F127M filter combination would work best with the least degeneracy in the color space (Figure \ref{f:hst:nb2:1col}): each subtype has a unique color, approximately equally spaced from the neighboring types. 
\begin{figure}
\includegraphics[width=0.5\textwidth]{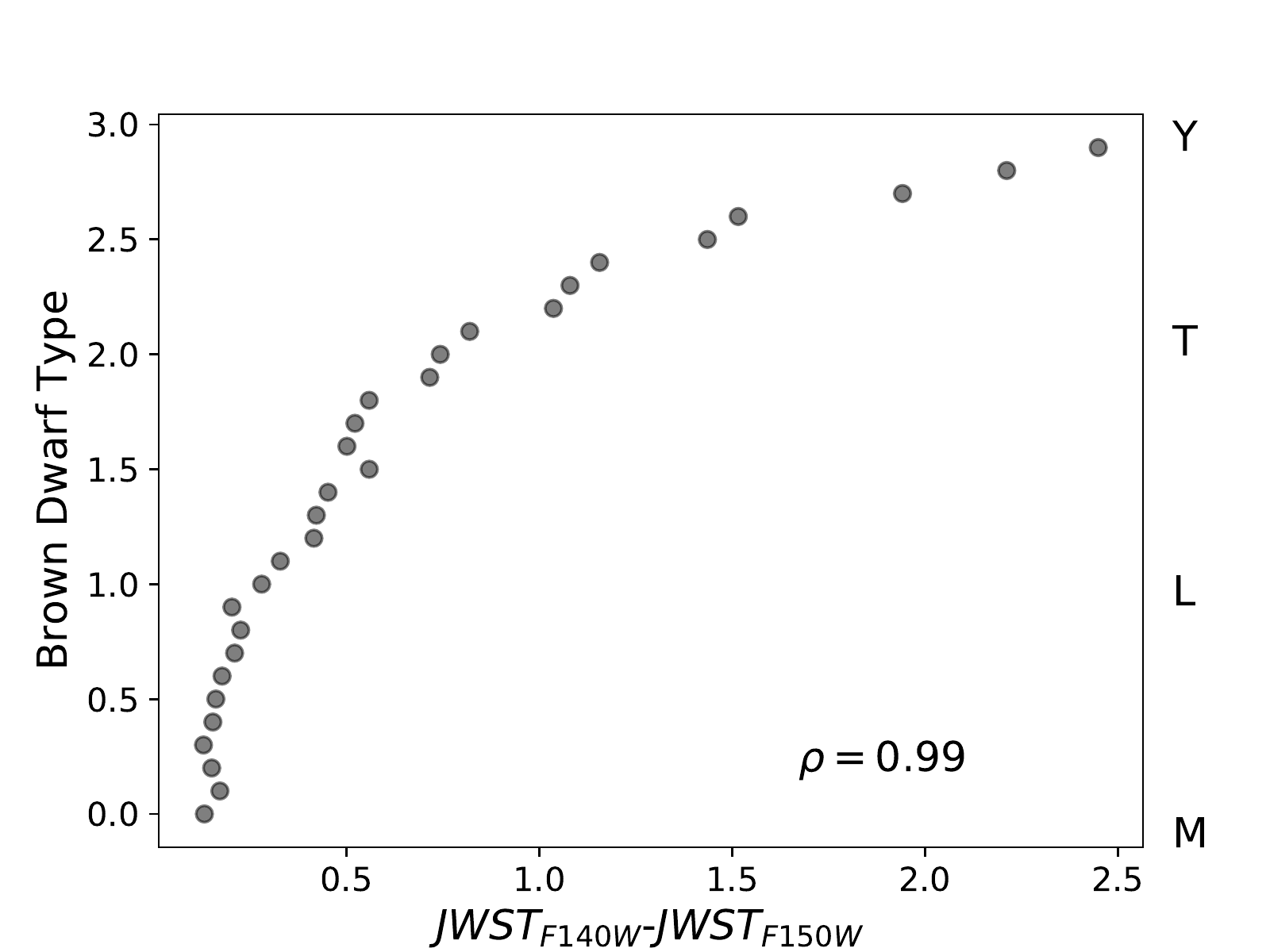}
\includegraphics[width=0.5\textwidth]{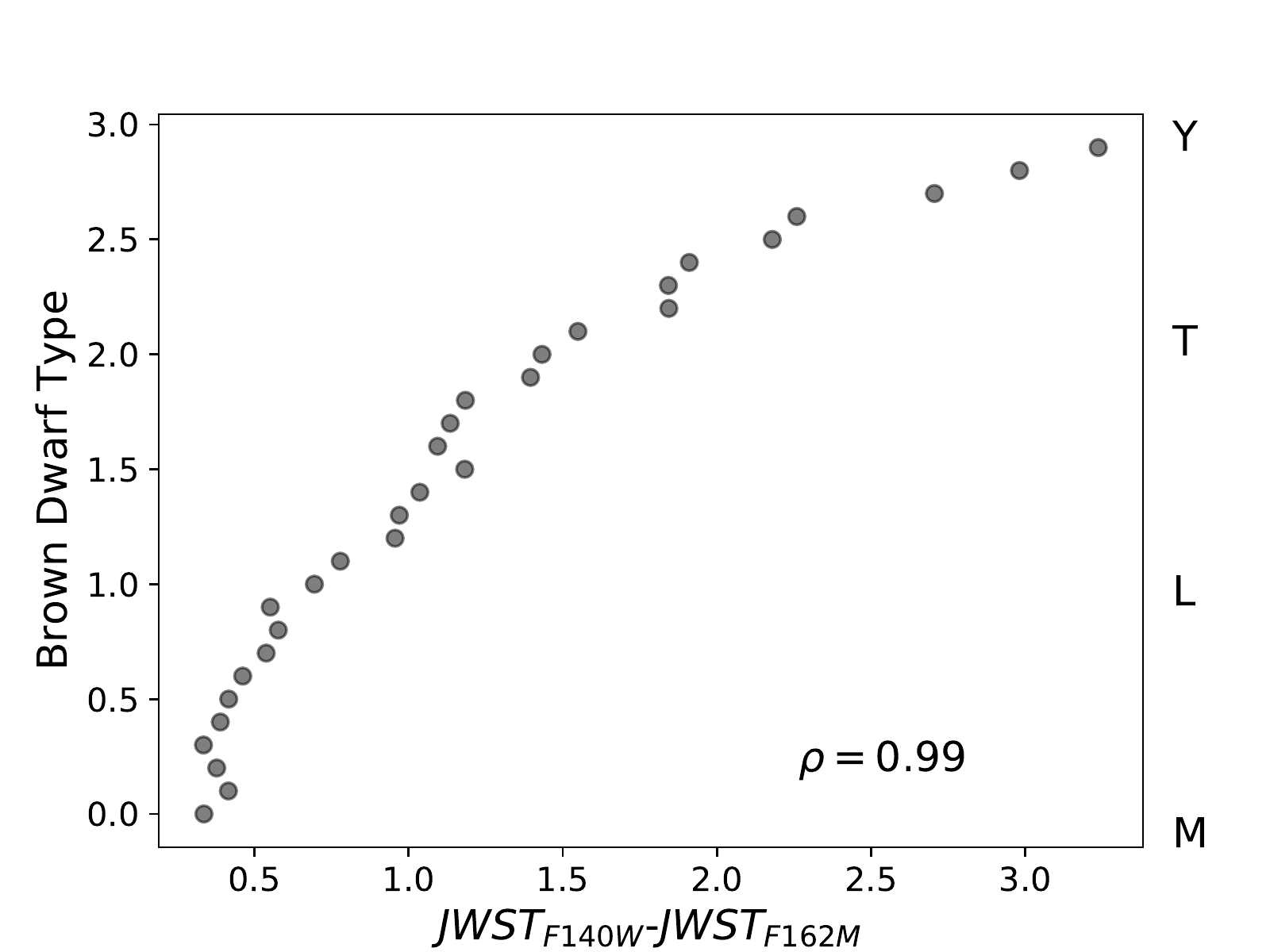}
\includegraphics[width=0.5\textwidth]{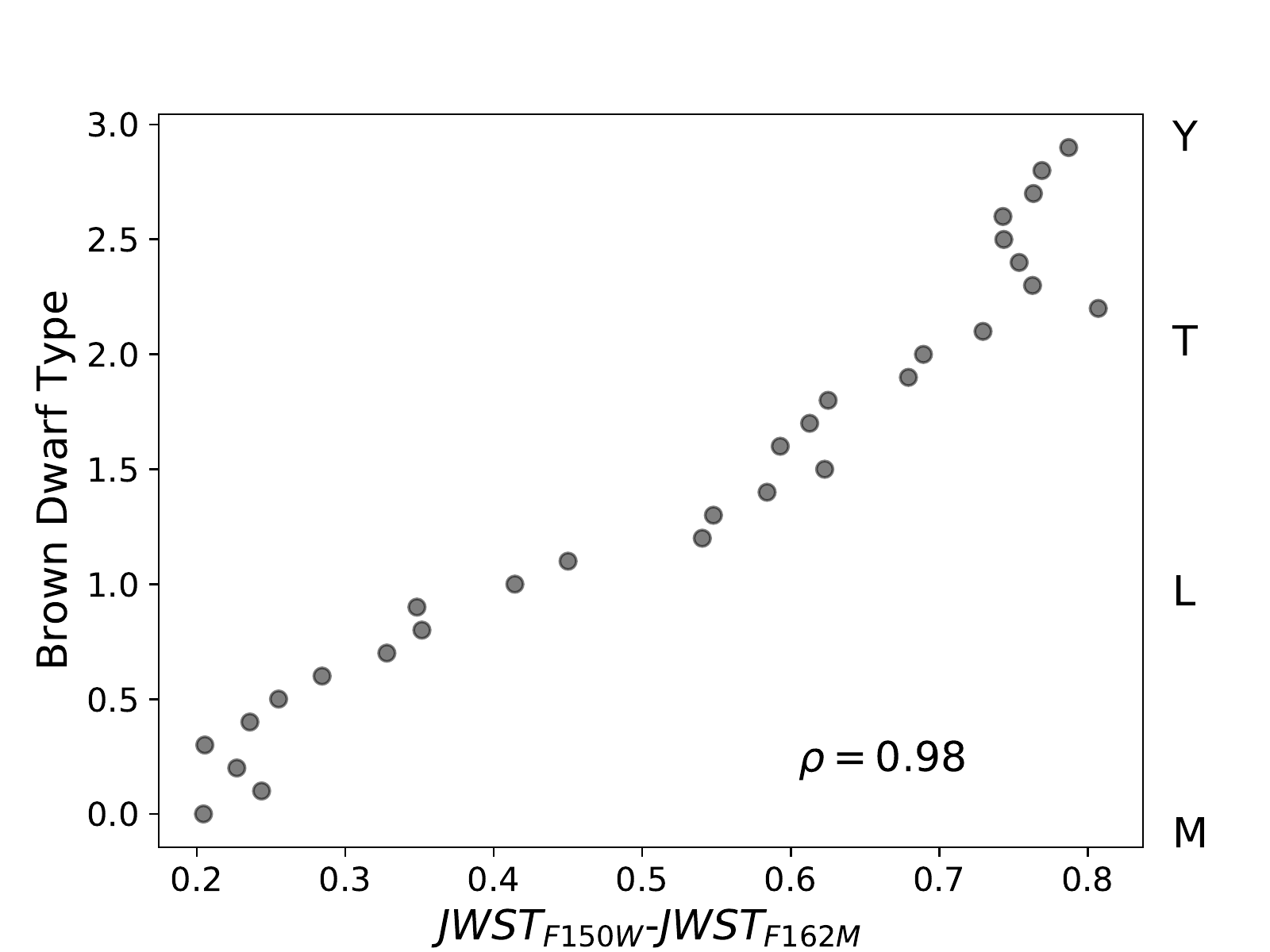}
\caption{\label{f:jwst:1col}Relation between three \emph{JWST} colors using the optimal filter combinations and the brown dwarf type and subtype for the standard stars in {\sc splat}. The F140W/F150W or the F140W/F162M filter combinations result unambiguous subtyping. the F150W/F162M combinations is degenerate in the Y-dwarf category.}
\end{figure}

The next mission to consider is \emph{JWST}, which a range of wide, medium, and narrow-abnd filters in this wavelength regime. Three of these stand out in their ability to separate out subtype. Figure \ref{f:jwst:1col} shows the brown dwarf type relation with three different color combinations using the F140M, F150W and F162M filters on board \emph{JWST/NIRCam}. The combination of these filters differentiate the different dwarf types very well from each other. The F140M-F150W and F140M-F162M filter combinations are equally good options to separate out red and brown dwarfs into different subtypes. The F150W-F162M combination  performs slightly poorer than either of the other two filter combinations with some degeneracy in the mid L-types (Figure \ref{f:jwst:1col}). 


We have already established that the \emph{Euclid} filters do not separate out broad brown dwarf types well and any correlation with subtypes is nonexistent.


For \emph{WFIRST}, only the Y106-F184 color ($\rho=0.7953$) correlates best with the brown dwarf (sub)type. Both filters are close to the best-performing \emph{HST} filter ($Y$-$I$) color. Like the \emph{HST} color-type relation in Figure \ref{f:hst:1col}, it is a relation that works best for M-dwarfs but not later types.


\subsection{Dwarf Subtyping with Combined Missions}
\label{s:1col:dual}

When considering combinations of \emph{HST} and \emph{JWST} filters, there is no  filter combination that works better than one of the combinations typically already available from \emph{HST} observations (see Table \ref{t:1col}).

While the combination of \emph{WFIRST} and \emph{Euclid} is somewhat better at identifying broad brown dwarf types (Figure \ref{f:euclid:wfirst:2col}), the filter combination that correlates best with type, a \emph{WFIRST} combination of Y106 and F184, still does not discriminate among types very well (Figure \ref{f:euclid:wfirst:2col} and Table \ref{t:2col}). \emph{WFIRST}'s mission concept is, at present, not an ideal follow-up instrument of brown dwarfs identified {\em with imaging}. \emph{WFIRST} would be markedly better at discerning between brown dwarf types with the addition of a single medium-band filter to \emph{WFIRST}'s filter selection. When a medium-band filter is combined with the wide W146 filter, for example, \emph{WFIRST} would become an effective mission for subtyping brown dwarfs.  
Arguably, this could also be done with \emph{WFIRST} grism capability if done to a similar limiting depth as the imaging surveys. The typing of substellar dwarfs is one of several reasons to add a medium-wide filter but brown dwarf characterization would be the argument for one centered on the key absorption features.

\section{Discussion}
\label{s:discussion}

In this work, we explore how well the current \emph{HST} filter set that is typically used in high-redshift observations (e.g., CANDELS, BoRG, XDF) could discriminate between both broad and subtypes of brown dwarfs. 
Our second goal is to explore to what extent future near-infrared imaging observatories can be used to type and subtype brown dwarfs.

Our results show that the current filter combinations for BoRG[z8] and BoRG[z9] observations are nearly as ideal for the discrimination between broad type as is practical with the \emph{HST} filters that are most commonly used for extragalactic work\footnote{BoRG has a single optical filter, either F606W (broad $V$) or F350LP, which do not help with broad dwarf typing but may be of use for subtyping M-dwarfs.}. 
The ability of the BoRG[z8] and BoRG[z9] filter combinations to distinguish broad substellar type. This holds promise to use the new BoRG[z9] observations to map the Milky Way population using the many separate sightlines. 

However, when all the medium- and narrow-band filters available in WFC3 are considered, good subtype separation can be achieved with the addition of just one narrow (e.g., F126N) or medium (F127M) filter to the already available F125W imaging. We note this is true for the standard stars and that the relation between color and subtype is much noisier for the whole sample in {\sc SPLAT}. This offers the opportunity to supplement the extragalactic fields with just single filter observations to accurately subtype the red and brown dwarfs in these fields. For the legacy fields with grism information, this is less useful, but for the pure-parallel surveys (e.g., BoRG) this is a viable path to discriminate between subtypes and thus test the possible relation between brown dwarf cooling and the scale height of their distribution in the Milky Way \citep[see][]{Ryan17}. 

\emph{JWST} will be the most suitable upcoming mission for distinguishing brown dwarf types in terms of discretionary power, not just between broad types but with the clear potential to map colors to subtypes with reasonable accuracy. 

We note that the high-redshift imaging surveys with \emph{JWST} will not automatically be able to identify dwarfs, but the filter combinations that would make this possible are now known and could potentially be included in such surveys (e.g., as part of pre-imaging to avoid targeting brown dwarfs for follow-up spectroscopy for example).

\emph{Euclid} will be extremely useful in mapping the broad dwarf class of the Milky Way as a whole, but typing, let alone subtyping, these stellar objects will be challenging without spectroscopy (e.g., grism observations) or ancillary data (ground-based optical observations of M-dwarfs, for example). 

\emph{WFIRST} performs better at discriminating broad dwarf types, but lacks a medium- or narrow-band filter that would be capable of discriminating among subtypes. We therefore argue strongly in favor of adding a medium-band filter in the $H$-band to the \emph{WFIRST} filter wheel. 
While dwarf star and substellar programs alone may not constitute a strong enough science case to justify a medium band filter, Solar System science has also expressed an interest in adding a $Ks$ broad-band filter \citep{Holler17}. Additionally, the addition of such a filter would strengthen the selection of high-redshift candidates (of order 10k objects at a given redshift) by weeding out the M/L/T dwarfs (an order or so more objects, depending on depth and target redshift).
If a medium-band filter in the broad $H$-band range is considered a valuable addition to, for example, planetary science and other \emph{WFIRST} mission objectives, the project should seriously consider adding one to the filter wheel.

Combining the ubiquitous \emph{Euclid} observations (all-sky) with \emph{WFIRST} observations improves the identifications of broad brown dwarf categories (M/L/T), but does not succeed at subtyping.
With a need to map the initial mass function of stars galaxy-wide \citep{El-Badry17}, a medium band filter on \emph{WFIRST} would fill the gap of knowledge between any \emph{JWST} observations sampling the stellar content of nearby galaxies and \emph{WFIRST} observations constraining their full stellar population.

\section{Concluding Remarks}
\label{s:conclusions}

We group our conclusions by the narrow-field (\emph{HST} and \emph{JWST}) and the wide-field (\emph{Euclid} and \emph{WFIRST}) observatories:

\begin{itemize}
\item The \emph{HST} filters used thus far for high-redshift searches (e.g., CANDELS and BoRG) are close to optimal to identify broad red or brown dwarf type (Figure \ref{f:hst:2col} and Table \ref{t:2col}). 
\item With the addition of medium and narrow-band filters not commonly used for extragalactic surveys, a good separation of subtypes can be achieved using the F127M and F125W filters that are most commonly used for extragalactic surveys (Figure \ref{f:hst2:2col:med} and Table \ref{t:1col}). 
\item The combination of three \emph{JWST} filters (F140M, F150W and F162M) split both the broad and subtypes of brown dwarfs (Figure \ref{f:jwst:2col}).
\item \emph{JWST} F140M and F150W is the optimal combination of filters to subtype brown dwarfs (Figure \ref{f:jwst:1col}). Alternatively, the combination of F150W and F162M works almost as well (Figure \ref{f:jwst:1col} and Table \ref{t:1col}).
\item \emph{Euclid} alone performs the poorest of all four NIR observatories in discriminating among M-, L-, and T-type brown dwarfs (Figure \ref{f:euclid:2col}). However, it groups brown dwarfs in a clear part of color-space for follow-up observations (to target or avoid).
\item The \emph{WFIRST} filters perform similarly to \emph{Euclid}, with the optimal combination W146, H158, F062 separating broad brown dwarf types but not able to discriminate between subtypes (Figure \ref{f:wfirst:2col}).
\item The combination of \emph{Euclid} and \emph{WFIRST}, using \emph{WFIRST}'s W146 and F062 filters and \emph{Euclid}'s $Y$-band filter, allows for a much better discrimination between broad brown dwarf categories (Figure \ref{f:euclid:wfirst:2col}). 
\item Subsequent subtyping with the combination of \emph{Euclid} and \emph{WFIRST} observations remains uncertain due to the lack of medium or narrow-band filters in this wavelength range (Figure \ref{f:euclid:wfirst:2col}).

\end{itemize}
\section*{Acknowledgements}
 
The authors would like to thank the anonymous referee for the constructive and thoughtful critique of the earlier draft.  
This research made use of {\sc Astropy}, a community-developed core Python package for Astronomy \citep{Astropy-Collaboration13a}. This research made use of {\sc Matplotlib}, a Python library for publication quality graphics \citep{Hunter07}. {\sc PyRAF} is a product of the Space Telescope Science Institute, which is operated by AURA for NASA. This research made use of both {\sc SciPy} \citep{scipy} and {\sc SPLAT} \citep{Burgasser17}. {\sc SPLAT} is an experimental, collaborative project of research students in Adam Burgasser's UCSD Cool Star Lab, aimed at teaching students how to do research by building their own analysis tools. Contributors to {\sc SPLAT} have included Christian Aganze, Jessica Birky, Daniella Bardalez Gagliuffi, Adam Burgasser (PI), Caleb Choban, Andrew Davis, Ivanna Escala, Aishwarya Iyer, Yuhui Jin, Mike Lopez, Alex Mendez, Gretel Mercado, Elizabeth Moreno Hilario, Johnny Parra, Maitrayee Sahi, Adrian Suarez, Melisa Tallis, Tomoki Tamiya, Chris Theissen and Russell van Linge. This project is supported by the National Aeronautics and Space Administration under Grant No. NNX15AI75G.

\bibliographystyle{aa} 
\bibliography{Bibliography}
\clearpage
\newpage

\appendix

\begin{table*}
\caption{The apparent magnitudes of all {\sc splat} sources computed in the \emph{HST} filters. Full tables for all four missions considered available online with this publication.}
\begin{center}
\begin{tabular}{lllllllllllllllllllllllllllllllllllllll}
\hline
Number & index type & unc & st. type & unc &  f814w & err & f098m & err & f105w & err & f125w & err & f140w & err & f160w & err & f110w & err & f126n & err & f127m & err & f128n & err & f130n & err & f132n & err & f139m & err & f153m & err & f164n & err & f167n & err\\
\hline
\hline
\hline
\hline
10209 & 2.45 & 0.251874 & 2.4 & 0.5 & 29.372312 & 0.737955 & 26.388198 & 0.019829 & 26.040271 & 0.017670 & 25.151282 & 0.010545 & 24.976077 & 0.014932 & 25.012600 & 0.018229 & 25.586065 & 0.014592 & 24.407990 & 0.004786 & 24.377040 & 0.005564 & 24.231875 & 0.006726 & 24.371324 & 0.007122 & 24.599358 & 0.017207 & 26.993115 & 0.094235 & 24.726078 & 0.016803 & 24.574274\\
11855 & 1.10 & 0.556161 & 0.9 & 0.5 & 26.141540 & 0.047187 & 24.378857 & 0.008233 & 23.993578 & 0.005929 & 23.258797 & 0.005871 & 23.019959 & 0.005195 & 22.749328 & 0.004908 & 23.686045 & 0.006147 & 23.148405 & 0.003815 & 23.050387 & 0.003717 & 22.917331 & 0.004465 & 22.957785 & 0.004144 & 22.910851 & 0.004539 & 23.128230 & 0.010952 & 22.732345 & 0.005327 & 22.361824\\
10144 & 1.50 & 1.446706 & 0.7 & 0.5 & -11.486000 & 0.116757 & -12.871494 & 0.021930 & -13.122074 & 0.017582 & -13.691276 & 0.014701 & -13.881244 & 0.015448 & -14.066726 & 0.019746 & -13.352015 & 0.018002 & -13.769097 & 0.007117 & -13.833633 & 0.011190 & -13.917844 & 0.018427 & -13.931957 & 0.008718 & -13.933358 & 0.015810 & -13.858432 & 0.035609 & -14.064632 & 0.019727 & -14.364993\\
11982 & 1.45 & 0.782517 & 0.8 & 0.5 & 28.966369 & 0.249016 & 27.058639 & 0.027124 & 26.666028 & 0.023796 & 25.942172 & 0.017572 & 25.812199 & 0.028278 & 25.618098 & 0.027475 & 26.363584 & 0.019801 & 25.731165 & 0.013378 & 25.606229 & 0.013156 & 25.456794 & 0.017095 & 25.504963 & 0.014354 & 25.515775 & 0.013234 & 26.200514 & 0.041865 & 25.630973 & 0.036443 & 25.036492\\
10421 & 1.70 & 1.172794 & 1.5 & 0.5 & 28.005479 & 1.083947 & 26.099946 & 0.162242 & 25.620674 & 0.092142 & 24.727244 & 0.060927 & 24.404745 & 0.053707 & 23.995905 & 0.066472 & 25.219328 & 0.080270 & 24.521394 & 0.030904 & 24.399724 & 0.048754 & 24.249587 & 0.061165 & 24.282875 & 0.028959 & 24.251556 & 0.086233 & 24.735634 & 0.079951 & 23.945005 & 0.060995 & 23.451594\\
10247 & 1.55 & 0.550069 & 1.5 & 0.5 & 26.594247 & 0.063245 & 24.575206 & 0.007908 & 24.030100 & 0.006780 & 23.055668 & 0.008470 & 22.736875 & 0.007388 & 22.343638 & 0.004645 & 23.582249 & 0.008403 & 22.880789 & 0.003389 & 22.722187 & 0.002531 & 22.541876 & 0.001526 & 22.595052 & 0.003459 & 22.532386 & 0.003934 & 23.103446 & 0.032210 & 22.293263 & 0.003764 & 21.810253\\
10698 & 1.15 & 0.771520 & 0.8 & 0.5 & -10.935690 & 0.031929 & -12.632574 & 0.005225 & -12.965344 & 0.004880 & -13.609522 & 0.004166 & -13.786967 & 0.004354 & -13.995748 & 0.004867 & -13.225395 & 0.004635 & -13.723168 & 0.003293 & -13.826193 & 0.003169 & -13.961135 & 0.003625 & -13.912346 & 0.002976 & -13.938581 & 0.003013 & -13.614185 & 0.007859 & -14.004456 & 0.004954 & -14.386689\\
11141 & 1.40 & 0.605644 & 1.2 & 0.5 & -10.322486 & 0.198560 & -12.062434 & 0.029495 & -12.590032 & 0.017821 & -13.522434 & 0.009380 & -13.818142 & 0.014007 & -14.169966 & 0.012898 & -13.008769 & 0.014654 & -13.682035 & 0.008707 & -13.819253 & 0.009125 & -13.992386 & 0.010182 & -13.944113 & 0.008279 & -14.000426 & 0.007713 & -13.592730 & 0.017340 & -14.199573 & 0.011799 & -14.658427\\
\dots & \dots & \dots & \dots & \dots & \dots & \dots & \dots & \dots & \dots & \dots & \dots & \dots & \dots \dots & \dots & \dots & \dots & \dots & \dots & \dots \\
\hline
\end{tabular}
\end{center}
\label{t:hst:all:mag}
\end{table*}%

\begin{table*}
\caption{The apparent magnitudes of the standard stars in {\sc splat} computed in the \emph{HST} filters.  Full tables for all four missions considered available online with this publication.}
\begin{center}
\begin{tabular}{lllllllllllllllllllllllllllllllllll}
\hline
type & f814w & err & f098m & err & f105w & err & f125w & err & f140w & err & f160w & err & f110w & err & f126n & err & f127m & err & f128n & err & f130n & err & f132n & err & f139m & err & f153m & err & f164n & err & f167n & err\\
\hline
\hline
0.000000 & 15.899930 & 0.001543 & 15.391747 & 0.001118 & 15.261395 & 0.001002 & 14.869515 & 0.000965 & 14.626093 & 0.000908 & 14.375303 & 0.001027 & 15.105016 & 0.000860 & 14.830084 & 0.000818 & 14.783623 & 0.000793 & 14.700590 & 0.000830 & 14.754157 & 0.000894 & 14.733106 & 0.000851 & 14.586792 & 0.001423 & 14.369414 & 0.000989 & 14.163077 & 0.000947 & 14.194602 & 0.001113 & \\
0.100000 & 14.854688 & 0.001903 & 14.280802 & 0.001044 & 14.145593 & 0.000972 & 13.773502 & 0.000825 & 13.553462 & 0.000915 & 13.308419 & 0.001013 & 14.000394 & 0.000991 & 13.707886 & 0.000826 & 13.648148 & 0.000755 & 13.545283 & 0.000861 & 13.623641 & 0.000800 & 13.608124 & 0.000885 & 13.631821 & 0.001444 & 13.292160 & 0.000886 & 13.078941 & 0.000978 & 13.102232 & 0.000981 & \\
0.200000 & 15.702178 & 0.009927 & 15.042564 & 0.006830 & 14.894865 & 0.006286 & 14.479290 & 0.006014 & 14.231508 & 0.004051 & 13.974077 & 0.001843 & 14.729068 & 0.005834 & 14.429890 & 0.004371 & 14.368598 & 0.004710 & 14.256646 & 0.004087 & 14.345302 & 0.003254 & 14.318873 & 0.003554 & 14.236413 & 0.010730 & 13.966399 & 0.001287 & 13.752624 & 0.000890 & 13.777591 & 0.001064 & \\
0.300000 & 14.611093 & 0.002699 & 13.882720 & 0.001286 & 13.732808 & 0.001237 & 13.326169 & 0.001144 & 13.107942 & 0.000999 & 12.882308 & 0.000879 & 13.573269 & 0.000882 & 13.273774 & 0.000862 & 13.229219 & 0.000912 & 13.149576 & 0.000855 & 13.201040 & 0.000657 & 13.179915 & 0.000841 & 13.094307 & 0.001748 & 12.878076 & 0.000817 & 12.674388 & 0.000873 & 12.693066 & 0.000739 & \\
0.400000 & 16.368592 & 0.003457 & 15.462166 & 0.001011 & 15.279736 & 0.000793 & 14.813992 & 0.000642 & 14.581105 & 0.000616 & 14.341436 & 0.000561 & 15.094559 & 0.000832 & 14.750009 & 0.000475 & 14.695918 & 0.000561 & 14.610911 & 0.000505 & 14.653582 & 0.000494 & 14.625309 & 0.000513 & 14.595311 & 0.001239 & 14.331436 & 0.000522 & 14.110528 & 0.000545 & 14.118771 & 0.000549 & \\
0.500000 & 17.839593 & 0.024412 & 16.780626 & 0.006790 & 16.577970 & 0.005355 & 16.090950 & 0.003326 & 15.881036 & 0.003220 & 15.662715 & 0.002252 & 16.384756 & 0.004814 & 16.012595 & 0.002400 & 15.954698 & 0.002661 & 15.861983 & 0.002482 & 15.904212 & 0.002485 & 15.875232 & 0.002724 & 15.922700 & 0.004876 & 15.652497 & 0.002480 & 15.406575 & 0.002257 & 15.401854 & 0.002174 & \\
0.600000 & 18.462491 & 0.011128 & 17.226143 & 0.006864 & 16.990912 & 0.007171 & 16.465313 & 0.005783 & 16.263692 & 0.004715 & 16.046787 & 0.002890 & 16.782723 & 0.007193 & 16.371868 & 0.004267 & 16.303667 & 0.004446 & 16.190199 & 0.005202 & 16.252163 & 0.003926 & 16.230800 & 0.003304 & 16.318688 & 0.008149 & 16.047508 & 0.002219 & 15.756212 & 0.002933 & 15.759329 & 0.003291 & \\
0.700000 & 19.690574 & 0.011464 & 18.236665 & 0.002748 & 17.969519 & 0.002390 & 17.415236 & 0.002319 & 17.245658 & 0.002314 & 17.053054 & 0.001763 & 17.751195 & 0.002383 & 17.290638 & 0.001944 & 17.220826 & 0.001506 & 17.110254 & 0.001818 & 17.158068 & 0.001576 & 17.143415 & 0.001623 & 17.365278 & 0.004036 & 17.046260 & 0.002109 & 16.720733 & 0.001564 & 16.702244 & 0.001713 & \\
0.800000 & 20.130288 & 0.017328 & 18.405429 & 0.003826 & 18.070077 & 0.003053 & 17.412810 & 0.002437 & 17.215094 & 0.002582 & 16.994888 & 0.002360 & 17.802972 & 0.002592 & 17.298781 & 0.001991 & 17.214076 & 0.001981 & 17.093139 & 0.001820 & 17.131105 & 0.001721 & 17.098945 & 0.001781 & 17.333887 & 0.003808 & 16.983724 & 0.002007 & 16.639809 & 0.001650 & 16.621532 & 0.001750 & \\
0.900000 & 22.103708 & 0.023542 & 20.268131 & 0.005144 & 19.861593 & 0.004005 & 19.096285 & 0.003403 & 18.877246 & 0.003753 & 18.646106 & 0.003500 & 19.539159 & 0.003206 & 19.010867 & 0.002669 & 18.921996 & 0.002929 & 18.807569 & 0.002711 & 18.821771 & 0.002674 & 18.764090 & 0.002541 & 18.894676 & 0.005295 & 18.644729 & 0.003201 & 18.279693 & 0.003100 & 18.255598 & 0.003109 & \\
1.000000 & 24.744026 & 0.099897 & 22.886584 & 0.018988 & 22.446538 & 0.018206 & 21.652739 & 0.013395 & 21.416498 & 0.013475 & 21.144286 & 0.013363 & 22.111148 & 0.017162 & 21.558256 & 0.013655 & 21.419669 & 0.014155 & 21.269720 & 0.013574 & 21.308717 & 0.013770 & 21.271402 & 0.007237 & 21.581955 & 0.016629 & 21.131820 & 0.010241 & 20.752074 & 0.010630 & 20.724678 & 0.009306 & \\
1.100000 & -10.494208 & 0.056565 & -12.332495 & 0.010081 & -12.780326 & 0.006745 & -13.579649 & 0.004256 & -13.817763 & 0.004967 & -14.101498 & 0.004503 & -13.119392 & 0.005727 & -13.732771 & 0.003828 & -13.836052 & 0.003597 & -13.986955 & 0.004018 & -13.936387 & 0.003849 & -13.984065 & 0.003181 & -13.607015 & 0.006016 & -14.141812 & 0.004562 & -14.533039 & 0.003968 & -14.549914 & 0.004152 & \\
1.200000 & -10.289958 & 0.091685 & -12.118031 & 0.013882 & -12.624098 & 0.009503 & -13.523631 & 0.006778 & -13.801444 & 0.007654 & -14.147596 & 0.006884 & -13.024292 & 0.007977 & -13.718354 & 0.004889 & -13.830464 & 0.005378 & -13.989410 & 0.004960 & -13.943684 & 0.004197 & -13.998707 & 0.004810 & -13.524781 & 0.015841 & -14.173500 & 0.005629 & -14.667099 & 0.004756 & -14.691105 & 0.004243 & \\
1.300000 & -10.177335 & 0.186718 & -12.037084 & 0.022890 & -12.560805 & 0.017159 & -13.494929 & 0.009837 & -13.774356 & 0.012136 & -14.116088 & 0.011859 & -12.983060 & 0.011522 & -13.719778 & 0.006404 & -13.826311 & 0.008380 & -13.981797 & 0.007555 & -13.945934 & 0.006973 & -13.972631 & 0.008361 & -13.513218 & 0.016063 & -14.133375 & 0.013291 & -14.642582 & 0.008693 & -14.681513 & 0.009229 & \\
1.400000 & 26.296216 & 0.228449 & 24.328822 & 0.033099 & 23.810042 & 0.022603 & 22.881367 & 0.015661 & 22.623482 & 0.016774 & 22.287825 & 0.019745 & 23.391235 & 0.021655 & 22.659144 & 0.009118 & 22.546164 & 0.010798 & 22.378866 & 0.012281 & 22.435334 & 0.010497 & 22.389697 & 0.010421 & 22.903040 & 0.055254 & 22.272367 & 0.017188 & 21.725578 & 0.011805 & 21.689883 & 0.013453 & \\
1.500000 & 27.106031 & 0.878145 & 25.151953 & 0.106404 & 24.650891 & 0.082973 & 23.728704 & 0.042004 & 23.415076 & 0.036829 & 23.008029 & 0.039664 & 24.231876 & 0.050880 & 23.470363 & 0.029227 & 23.369098 & 0.029498 & 23.207328 & 0.034722 & 23.270314 & 0.026719 & 23.213718 & 0.038970 & 23.839673 & 0.062297 & 22.951460 & 0.036749 & 22.433557 & 0.034081 & 22.428289 & 0.033647 & \\
1.600000 & 29.001420 & 0.241383 & 27.025268 & 0.031174 & 26.540100 & 0.021834 & 25.594836 & 0.014624 & 25.255701 & 0.014802 & 24.844285 & 0.011932 & 26.103488 & 0.017385 & 25.374709 & 0.011522 & 25.257928 & 0.015131 & 25.104765 & 0.018181 & 25.148398 & 0.012580 & 25.078741 & 0.009620 & 25.587854 & 0.020240 & 24.804208 & 0.010279 & 24.302830 & 0.007837 & 24.269766 & 0.008438 & \\
1.700000 & 27.507458 & 0.451258 & 25.263335 & 0.040385 & 24.803500 & 0.026976 & 23.846489 & 0.017546 & 23.471683 & 0.018497 & 23.024781 & 0.017005 & 24.350617 & 0.021934 & 23.575645 & 0.012359 & 23.494574 & 0.013865 & 23.364886 & 0.015160 & 23.392163 & 0.012365 & 23.350688 & 0.011127 & 23.762839 & 0.034608 & 23.006614 & 0.014468 & 22.458487 & 0.010767 & 22.430675 & 0.010420 & \\
1.800000 & 27.110130 & 0.613350 & 24.879452 & 0.036599 & 24.466824 & 0.028260 & 23.554237 & 0.027429 & 23.177565 & 0.031327 & 22.718103 & 0.028558 & 24.037548 & 0.029152 & 23.258687 & 0.023045 & 23.193358 & 0.019714 & 23.092559 & 0.023687 & 23.109466 & 0.025067 & 23.087771 & 0.018796 & 23.504790 & 0.059749 & 22.690472 & 0.029055 & 22.142752 & 0.022424 & 22.122829 & 0.021579 & \\
1.900000 & 26.508975 & 0.354354 & 24.183902 & 0.025975 & 23.773832 & 0.015019 & 22.904929 & 0.024973 & 22.547602 & 0.022890 & 22.088120 & 0.013392 & 23.373041 & 0.022646 & 22.544341 & 0.009017 & 22.462936 & 0.007256 & 22.325836 & 0.009688 & 22.392331 & 0.005775 & 22.424173 & 0.011394 & 23.154937 & 0.154122 & 22.007046 & 0.011298 & 21.490820 & 0.004501 & 21.502957 & 0.006692 & \\
2.000000 & 27.009912 & 0.265701 & 24.632475 & 0.023881 & 24.238698 & 0.018373 & 23.391026 & 0.012787 & 23.114955 & 0.012633 & 22.762825 & 0.011700 & 23.845894 & 0.012924 & 22.968935 & 0.007790 & 22.913919 & 0.007579 & 22.803333 & 0.007163 & 22.834888 & 0.007287 & 22.864492 & 0.007241 & 23.865499 & 0.049167 & 22.660831 & 0.011524 & 22.164018 & 0.007926 & 22.189194 & 0.008374 & \\
2.100000 & 28.690396 & 0.923623 & 25.936288 & 0.069293 & 25.565081 & 0.045821 & 24.746977 & 0.029511 & 24.504619 & 0.035155 & 24.183535 & 0.035088 & 25.184462 & 0.036782 & 24.329196 & 0.019452 & 24.241748 & 0.018745 & 24.119494 & 0.019296 & 24.149339 & 0.018603 & 24.172130 & 0.017187 & 25.370212 & 0.123533 & 24.105119 & 0.037800 & 23.528799 & 0.022701 & 23.548109 & 0.023805 & \\
2.200000 & -9.207141 & 0.346215 & -11.947037 & 0.027084 & -12.332387 & 0.022844 & -13.215783 & 0.019505 & -13.455624 & 0.017961 & -13.756934 & 0.024540 & -12.762294 & 0.020081 & -13.761028 & 0.007704 & -13.832774 & 0.011954 & -13.967727 & 0.014547 & -13.896515 & 0.014923 & -13.811501 & 0.011333 & -12.208080 & 0.074376 & -13.846490 & 0.022130 & -14.435888 & 0.010848 & -14.396686 & 0.020042 & \\
2.300000 & 30.155773 & 1.000065 & 27.364090 & 0.040006 & 27.022633 & 0.032188 & 26.199667 & 0.026212 & 25.980341 & 0.025047 & 25.806879 & 0.028144 & 26.618906 & 0.026953 & 25.571468 & 0.012710 & 25.521290 & 0.014615 & 25.379409 & 0.015490 & 25.502703 & 0.012380 & 25.636102 & 0.013254 & 27.551760 & 0.150409 & 25.614419 & 0.025308 & 25.285149 & 0.021292 & 25.324547 & 0.025732 & \\
2.400000 & -8.694234 & 0.420600 & -11.698319 & 0.024193 & -12.109211 & 0.018304 & -13.059519 & 0.010441 & -13.255685 & 0.015690 & -13.300924 & 0.018336 & -12.591110 & 0.015084 & -13.726148 & 0.004460 & -13.785658 & 0.005676 & -13.924480 & 0.007166 & -13.833914 & 0.009277 & -13.696647 & 0.009550 & -11.545001 & 0.093878 & -13.547346 & 0.013853 & -13.733687 & 0.013750 & -13.635231 & 0.023131 & \\
2.500000 & -8.549112 & 0.347421 & -11.678520 & 0.018601 & -11.985034 & 0.014403 & -12.901239 & 0.011585 & -13.072898 & 0.014591 & -12.844708 & 0.024555 & -12.476895 & 0.012099 & -13.760193 & 0.006258 & -13.770058 & 0.004990 & -13.916881 & 0.005098 & -13.743727 & 0.002974 & -13.439327 & 0.018078 & -10.445431 & 0.120726 & -13.241676 & 0.017079 & -13.071657 & 0.015150 & -12.849973 & 0.032326 & \\
2.600000 & 29.991384 & 0.773039 & 27.071246 & 0.029344 & 26.793195 & 0.024424 & 25.799084 & 0.019202 & 25.595478 & 0.023286 & 25.905353 & 0.031616 & 26.229258 & 0.022578 & 24.858067 & 0.006808 & 24.846328 & 0.010674 & 24.684421 & 0.011652 & 24.892664 & 0.012939 & 25.205128 & 0.010705 & 28.434166 & 0.738226 & 25.471823 & 0.020860 & 25.876853 & 0.023218 & 26.028462 & 0.031771 & \\
2.700000 & 30.448999 & 1.103697 & 27.126489 & 0.044075 & 26.949189 & 0.036304 & 25.966605 & 0.020832 & 25.704727 & 0.030168 & 26.075032 & 0.051842 & 26.361753 & 0.028608 & 24.844237 & 0.011778 & 24.894955 & 0.009334 & 24.740085 & 0.007872 & 25.064345 & 0.009808 & 25.570628 & 0.011345 & 29.324037 & 0.724467 & 25.554014 & 0.031693 & 26.395775 & 0.077805 & 26.622422 & 0.098748 & \\
2.800000 & 29.948017 & 0.783433 & 26.908346 & 0.023648 & 26.759056 & 0.022226 & 25.703298 & 0.015412 & 25.350569 & 0.016513 & 25.629902 & 0.029105 & 26.103868 & 0.016178 & 24.407874 & 0.005125 & 24.533197 & 0.005388 & 24.405737 & 0.005818 & 24.833399 & 0.007985 & 25.443075 & 0.011160 & 30.124869 & 0.976734 & 25.094454 & 0.017494 & 26.414867 & 0.063849 & 26.627387 & 0.082270 & \\
2.900000 & 30.917000 & 0.881172 & 28.052596 & 0.042853 & 28.024125 & 0.045980 & 27.142792 & 0.036705 & 26.725816 & 0.037469 & 27.013187 & 0.066687 & 27.445632 & 0.031774 & 25.615928 & 0.009531 & 25.856703 & 0.011106 & 25.768459 & 0.012957 & 26.471303 & 0.018044 & 27.345420 & 0.034229 & 30.682969 & 1.400790 & 26.474543 & 0.038656 & 28.286133 & 0.251635 & 28.909125 & 0.589481 & \\

\hline
\end{tabular}
\end{center}
\label{t:hst:std:mag}
\end{table*}%

\begin{table*}
\caption{The top three Spearman rankings between type and subtype and two-colors using filter combinations onboard 
\emph{HST}, \emph{JWST}, and \emph{WFIRST}, using the standard stars in {\sc splat}.}
\begin{center}
\begin{tabular}{|c|c|}
\hline
Filter combination & Spearman Ranking ($\rho$) and validity (p-value)\\
\hline
\hline
\emph{HST} & \\
\hline
$I_{F814W}$--$Y_{F098M}$ \& $Y_{F098M}$--$JH_{F140W}$ & 0.8186 (0.00) \\ 
$I_{F814W}$--$Y_{F105W}$ \& $Y_{F105W}$--$JH_{F140W}$ & 0.8179 (0.00) \\ 
$I_{F814W}$--$J_{F125W}$ \& $J_{F125W}$--$JH_{F140W}$ & {\bf 0.8197} (0.00) \\  
\hline
\hline
\emph{JWST} & \\
\hline
${F070W}$--$ {F115W}$ \& $ {F115W}$--$ {F150W}$ & 0.8215 (0.00) \\ 
$ {F070W}$--$ {F115W}$ \& $ {F115W}$--$ {F162M}$ & 0.8581 (0.00) \\ 
$ {F140M}$--$ {F150W}$ \& $ {F150W}$--$ {F162M}$ & {\bf 0.9897} (0.00) \\  
\hline
\hline
\emph{WFIRST} & \\
\hline
$ {W146}$--$ {F062}$ \& ${F062}$--$ {H158}$ & 0.6807 (0.00) \\ 
$ {W146}$--$ {H158}$ \& $ {H158}$--$ {F062}$ & {\bf 0.6922} (0.00) \\  
$ {J129}$--$ {H158}$ \& $ {H158}$--$ {F062}$ & 0.6624 (0.00) \\ 
\hline

\end{tabular}
\end{center}
\label{t:2col}
\end{table*}%

\begin{table*}
\caption{The top three Spearman rankings between (sub)type and a single filter combination
onboard \emph{HST}, \emph{JWST}, or \emph{WFIRST}, 
and the \emph{HST}/\emph{JWST} and \emph{EUCLID}/\emph{WFIRST} combinations 
using {\em all} the stars in {\sc splat}.}
\begin{center}
\begin{tabular}{|c|c|}
\hline
Filter combination & Spearman Ranking ($\rho$)  \\
\hline
\hline
\emph{HST} (Broad Filters) & \\
\hline
$Y_{F098M}$--$J_{F125W}$ & 0.7686 (0.00) \\ 
$Y_{F098M}$--$JH_{F140W}$ & 0.7738 (0.00) \\ 
$Y_{F098M}$--$H_{F160W}$ & {\bf 0.7769} (0.00) \\  
\hline
\hline
\emph{HST} (broad/medium/narrow filters) & \\
\hline
$Y_{F098M}$--$H_{F160W}$ & 0.7784 (0.00) \\ 
$Y_{F098M}$--F110W & 0.7791 (0.00) \\ 
$Y_{F098M}$--F164N & {\bf 0.7817} (0.00) \\  
\hline
\hline
\emph{HST} (F125W+medium/narrow filters) & \\
\hline
$J_{F125W}$--F126N & 0.5436 (0.00) \\ 
$J_{F125W}$--F127M & 0.6877 (0.00) \\ 
$J_{F125W}$--F128N & 0.7226 (0.00) \\ 
$J_{F125W}$--F130N & 0.7208 (0.00) \\ 
$J_{F125W}$--F132N & 0.7152 (0.00) \\ 
$J_{F125W}$--F139M & 0.6792 (0.00) \\ 
\hline
\hline
\emph{JWST} & \\
\hline
$ {F0790W}$--$ {F115W}$ & 0.9911 (0.00) \\ 
$ {F140M}$--$ {F150W}$ & 0.9898 (0.00) \\ 
$ {F140M}$--$ {F162M}$ & {\bf 0.9933} (0.00) \\  
\hline
\hline
\emph{WFIRST} & \\
\hline
$WFIRST_{Y106}$--$WFIRST_{J129}$ & 0.7879 (0.00) \\ 
$WFIRST_{Y106}$--$WFIRST_{H158}$ & 0.7761 (0.00) \\ 
$WFIRST_{Y106}$--$WFIRST_{f184}$ & {\bf 0.7953} (0.00) \\ 
\hline
\hline
\emph{HST}/\emph{JWST} & \\
\hline 
$ {F140M}$--$ {F150W}$ & {\bf 0.6827} (0.00) \\  
$ {F140M}$--$ {F162M}$ & 0.6761 (0.00) \\ 
$ {F150W}$--$ {F162M}$ & 0.6634 (0.00) \\ 
\hline
\hline
\emph{EUCLID}/\emph{WFIRST} & \\
\hline 
$EUCLID_{H}$--$WFIRST_{Y106}$ & -0.7787 (0.00) \\ 
$WFIRST_{Y106}$--$WFIRST_{J129}$ & 0.7879 (0.00) \\ 
$WFIRST_{Y106}$--$WFIRST_{f184}$ & {\bf 0.7953} (0.00) \\ 
\hline
\end{tabular}
\end{center}
\label{t:1col}
\end{table*}%



\end{document}